\documentclass[12pt,a4paper]{article}
\pdfoutput=1
\usepackage{graphicx}
\usepackage{jheppub}
\usepackage{amsmath,amssymb,euscript,array,mathrsfs,epsfig}
\usepackage[small]{caption}
\usepackage{dsfont}

\def\ben
{\begin{equation}}
\def\een{\end{equation}}

\def\p{{\bf p}}

    \let\L=\Lambda
   
 \let\W=\mu

\def\W={\cal W}
\def\L ={\cal L}

\def\be{\begin{equation}}
\def\ee{\end{equation}}
\def\ba{\begin{array}}
\def\ea{\end{array}}

\def\dalemb#1#2{{\vbox{\hrule height .#2pt
        \hbox{\vrule width.#2pt height#1pt \kern#1pt
                \vrule width.#2pt}
        \hrule height.#2pt}}}

\newcommand{\bea}{\begin{eqnarray}}
\newcommand{\eea}{\end{eqnarray}}

\title{Roton-phonon excitations in Chern-Simons matter theory at finite density}
\author{S. Prem Kumar, Dibakar Roychowdhury and Stanislav Stratiev }
\affiliation{Department of Physics,\\
Swansea University,\\
Singleton Park, Swansea,\\
SA2 8PP, U.K.}
\emailAdd{s.p.kumar@swansea.ac.uk, dibakarphys@gmail.com, stanislavstratiev@gmail.com}
\abstract{We consider $SU(N)$ Chern-Simons theory coupled to a scalar field in the fundamental representation at strictly zero temperature and  finite chemical potential for  the global $U(1)_B$  particle number or flavour symmetry.  In the semiclassical regime we identify a Bose condensed ground state with a vacuum expectation value (VEV) for the scalar accompanied by noncommuting  background gauge field matrix VEVs. These matrices coincide with the  droplet ground state of the Abelian quantum  Hall matrix model.
The ground state spontaneously breaks $U(1)_B$  and Higgses the gauge group whilst preserving spatial rotations and a colour-flavour locked global  $U(1)$ symmetry.  We compute the perturbative spectrum of semiclassical fluctuations for the $SU(2)$ theory and show the existence of a single massless state with a linear phonon dispersion relation and a roton minimum (and maximum) determining the Landau critical superfluid velocity. For the massless scalar theory with vanishing self interactions, the semiclassical dispersion relations and location of roton extrema take on  universal forms.}
\begin{document}
\maketitle
\flushbottom
\section{Introduction}
Relativistic field theories in three dimensions consisting of Chern-Simons gauge fields coupled to matter have been conjectured to enjoy a Bose-Fermi/level-rank duality symmetry \cite{Giombi:2011kc}. Mounting evidence for the conjecture has appeared in various forms. These include detailed aspects of  correlators \cite{Giombi:2011kc, Aharony:2011jz, Maldacena:2011jn, Maldacena:2012sf, Aharony:2012nh} and S-matrices \cite{Jain:2014nza, Dandekar:2014era} in large-$N$ vector models coupled to Chern-Simons gauge fields in the 't Hooft limit when the theory  becomes exactly solvable. 
Further, the large-$N$ thermal partition functions have been shown to exhibit Bose-Fermi duality as the 't Hooft coupling is varied \cite{Giombi:2011kc, Aharony:2012ns,Jain:2013py, Jain:2013gza, Takimi:2013zca}.  A crucial role in this is played by the nontrivial eigenvalue distributions of the holonomy matrix around the Euclidean thermal circle, and the duality manifests itself in various phases characterized by the large-$N$ eigenvalue distributions. The  finite $N$ versions of the duality can be precisely formulated \cite{Aharony:2015mjs}, and include an intricate web of abelian dualities \cite{Seiberg:2016gmd, Karch:2016sxi, Murugan:2016zal} with particle-vortex duality as one of its strands.

 In this work, motivated by the goal of understanding the manifestations of Bose-Fermi duality at finite density, we  study the zero temperature ground states of a fundamental scalar coupled to Chern-Simons gauge fields in the presence of a chemical potential for particle number. In particular, we will be mainly interested in finite density ground states in the (semi-)classical limit which spontaneously break the global $U(1)_B$ particle number symmetry. This is a subtle issue in 2+1 dimensions, as any finite temperature will result in thermal fluctuations that, by the Coleman-Mermin-Wagner theorem \cite{Coleman:1973ci, Mermin:1966fe}, can  destroy long-range order. 
 For this reason, in this paper we limit ourselves to the system at zero temperature. 
 At  non-zero temperature and finite density in the absence of condensates,  exact results at large-$N$ for Chern-Simons theory with a fundamental fermion \cite{Geracie:2015drf, Gur-Ari:2016xff} show nontrivial agreement at strong 't Hooft coupling\footnote{The 't Hooft coupling $\lambda$ is defined in the limit $N,k\to\infty$  ($k$ is the Chern-Simons level) where $\lambda \,\equiv\,\frac{N}{k}$, ranging between $0$ and $1$. } with the weakly interacting bosonic counterpart.
 
 In our analysis of the Chern-Simons-scalar system we  assume a classical limit i.e.  the Chern-Simons level $k$ is large (but finite), and any other scalar self-couplings taken to be  suitably weak so that the semiclassical description applies.  Our main findings are summarized below:
 \begin{itemize}
 \item{We find that the theory with $SU(N)$ gauge group, Chern-Simons level $k$ and non-zero chemical potential for particle number, exhibits a zero temperature ground state where  the scalar field condenses and all gauge fields acquire noncommuting background expectation values.  This ground state breaks the $SU(N)$ gauge symmetry completely and spontaneously breaks the global $U(1)_B$ particle number symmetry. While spatial rotations act nontrivially on the background gauge potentials, they  can be undone by a $U(1)_C$ subgroup of global $SU(N)$ transformations.  Thus gauge invariant operators acquire rotationally  invariant expectation values. The scalar VEV itself is left invariant by a combination of the flavour $U(1)_B$ and global colour $U(1)_C$ rotations. }
 \item{For the $SU(2)$ theory, assuming $k\gg 1$, we obtain the spectrum of physical fluctuations and their dispersion relations in the Bose condensed ground state. The fluctuation spectrum exhibits a massless phonon mode with linear dispersion relation 
 for the frequency $\omega\sim c_s |{\bf k}|$, for low spatial momenta ${\bf k}$, accompanied by a local maximum and a roton minimum at some finite spatial momentum.  Roton-maxon excitations are well known within the context of superfluidity in ${}^4{\rm He}$  \cite{Landau:1941vsj, schmitt} and explain various physical characteristics such as heat capacity and the superfluid critical velocity. The roton minimum, for instance, lowers the superfluid critical velocity to below the speed of sound, as can be understood by applying the Landau criterion \cite{Landau:1941vsj, schmitt}. In the context of this paper, we understand the appearance of the roton minimum as a consequence of level crossing of states. In the strict limit $k\to \infty$ when the Chern-Simons fields decouple, the interacting scalar theory has a Bose condensate with two gapless excitations at zero momentum, one with quadratic and the other with a linear dispersion relation. At  large but finite $k$, the former acquires a gap at zero momentum, and the putative intersection between the linear and quadratic dispersion curves is replaced by a roton-maxon pair in the diagonalized spectrum. The background VEVs for the gauge fields are directly responsible for these features. Roton-like excitations with very similar origin i.e. constant background gauge fields have been identified in Yang-Mills-Higgs systems at finite density in 3+1 dimensions \cite{Gusynin:2003yu}.
 
 We find that the roton minimum in the phonon dispersion relation persists in the free scalar theory coupled to Chern-Simons gauge fields (at large $k$). In this case the only dimensionful scale is provided by the chemical potential which can be rescaled to unity and the resulting spectra and dispersion relations acquire a universal form.  }
 \item{For the general $SU(N)$ case an interesting picture emerges. The $N\times N$ matrices of VEVs for the Chern-Simons gauge fields provide  finite dimensional versions of harmonic oscillator  creation and annihilation operators. In particular, they can be viewed as the noncommuting coordinates of $N$ particles in a disc of fixed radius. The same matrices have been used to describe the ground state of the quantum Hall droplet \cite{Polychronakos:2001mi, Susskind:2001fb}. Fluctuations about the finite density ground state  may thus be viewed as fluctuations of this droplet (in configuration space), carrying spatial momentum and frequency.}
  \end{itemize}
 The zero temperature finite density properties of the Chern-Simons-scalar system present a range of physical phenomena interesting in their own right. Importantly, they provide predictions for the fermionic dual.  The $SU(N)_k$  theory with a fundamental scalar (and level $k$) is level-rank dual to the $U(k-N)_{-k, -N}$ theory\footnote{The two subscripts denote the Chern-Simons levels of the $SU(k-N)$ and $U(1)$ factors of the gauge group respectively.} with a fundamental fermion \cite{Aharony:2015mjs}. In particular, the free scalar coupled to Chern-Simons fields is dual to the Chern-Simons plus critical fermion theory \cite{Minwalla:2015sca}. It is clearly of great interest to understand whether features of the spectrum of the weakly coupled scalar system can be understood from the conjectured fermionic dual at strong coupling.
 
 This paper is organized as follows. In section \ref{sec2} we study the Bose condensed ground state of the $SU(2)$ system in the classical limit. In section \ref{sec3} we find the spectrum of quadratic fluctuations after gauge fixing, and identify the phonon-roton branch for different regimes of parameters. Section \ref{sec4} is devoted to the generalization of the classical vacuum structure to general $N>2$.  Finally we outline a number of questions for future study in section \ref{sec5}.
 
  \section{The $SU(2)_k$ theory}
  \label{sec2}
We consider Chern-Simons theory with $SU(2)$ gauge group and one scalar flavour transforming in the fundamental representation. Working with an 
 anti-hermitean gauge potential $A_\mu$, 
\be
A_\mu\,=\,A_{\mu}^{(a)}\,t^a\,,\qquad t^a\,\equiv\,\frac{i}{2}\sigma^a\,,\qquad a=1,2,3\,,
\ee
where $\left\{\sigma^a \right\}$ are the Pauli matrices and $\{A_\mu^{(a)}\}$ are real valued fields,  the Chern-Simons action with (quantized) level $k$ is then,
\be
S_{\rm CS}\,=\,\frac{k}{4\pi}\int d^3x\,\epsilon^{\mu\nu\rho}\,{\rm Tr}\left(A_\mu\partial_\nu A_\rho\,+\,\frac{2}{3}A_\mu A_\nu A_\rho\right)\,.\label{E3}
\ee
This is the action for both Euclidean $(+++)$ and Lorentzian $(-++)$ signatures. The Wick rotation  from Lorentzian to Euclidean is implemented by the replacement $t\to -i\tau$ and $A_0 \to i A_0$, which together leave $S_{\rm CS}$ invariant.  In Lorentzian signature, the complete action involving Chern-Simons and matter fields has the general form,
\be
S\,=\,S_{\rm matter}\,+\, S_{\rm CS}\,,
\ee
where, in Lorentzian signature $(-++)$, for a scalar $\Phi$ transforming in the fundamental representation of $SU(2)$,
\bea
&&\,S_{\rm matter}\,=\,-\int d^3x\left(\left(D_\mu\Phi\right)^\dagger\left(D^\mu\Phi\right)\,+\,V(\Phi^\dagger \Phi)\right)\,,\label{E4}\\\nonumber\\\nonumber
&& D_\mu\,\equiv\,\partial_\mu\,+\,A_\mu\,.
\eea
The theory possesses a {\em global} $U(1)$ symmetry which we refer to as  ``baryon number" or $U(1)_B$,
\be
U(1)_B:\qquad \Phi\,\to\,e^{i\vartheta}\,\Phi\,,
\ee
generated by a phase rotation of $\Phi$. The corresponding conserved current is 
\be
j^\mu_B\,=\,i\,\left[\left(D^\mu\Phi\right)^\dagger \Phi\,-\,\Phi^\dagger D^\mu\Phi\right]\,.
\ee
The chemical potential $\mu_B$ is a Lagrange multiplier for the $U(1)_B$ charge. In Lorentzian signature, it therefore appears in the Lagrangian as a time component for a background $U(1)_B$ gauge field:
\be
D_\nu\,\to\, D_\nu\,+\,i\mu_B \,\delta_{\nu,0}\,.
\ee
\subsection{Classical ground states with $\mu_B\neq 0$}
The coupling of the Chern-Simons fields to the matter sector is controlled by $1/\sqrt{k}$\,\footnote{This can be understood via the rescaling $A_\mu \to A_\mu/\sqrt{k}$, following which the Chern-Simons action is order $1$ in the large $k$ limit.}. 
In the limit $k\to \infty$, the scalar field $\Phi$ with $\mu_B\neq 0$ has the potential:
\be
V_{\rm scalar}(\mu_B,\,k\to\infty)\,=\,V(\Phi^\dagger\Phi)\,-\,\mu_B^2\,\Phi^\dagger\Phi\,.
\ee
As usual, the effective negative mass squared due to the chemical potential drives the system to form a Bose condensate for large enough $\mu_B$. The tree level 3D scalar potential (at $\mu_B=0$) can be taken to  be of the form,
\be
V(\Phi^\dagger\Phi)\,=\,m^2\,\Phi^\dagger\Phi\,+\,g_4(\Phi^\dagger\Phi)^2\,+\,g_6\,(\Phi^\dagger\Phi)^3\,,
\ee
where we have allowed for relevant and marginal operators in the scalar potential.
Assuming that the ground state of the theory with $\mu_B\neq 0$ is static and translation invariant, we look for vacuum solutions with all terms involving derivatives being set to zero.  Anticipating a scalar condensate at the classical level\footnote{The analysis will remain purely classical and at zero temperature at this stage. At finite temperature, we know that quantum thermal fluctuations in 2+1 dimensions preclude symmetry breaking of continuous global symmetries.}, we can always choose gauge rotations to take the VEV to be real and of the form, 
\be
\langle\Phi\rangle\,=\,\left(
\begin{matrix}
 0 \\ v
\end{matrix}
\right)\qquad v\in {\mathbb R}\,.
\ee
We then collectively view all non-derivative terms as potential energy contributions:
\bea
V_{\rm CS}\,+\,V_{\rm scalar}&&=\,-\frac{k}{4\pi}\epsilon^{\mu\nu\rho}A_\mu^{(1)}\,A_\nu^{(2)}\,A_\rho^{(3)}\,-\, \frac{v^2}{4}\left[\left(A_0^{(1)}\right)^2 \,+\,\left(A_0^{(2)}\right)^2\,+\,\left(A_0^{(3)}\, -\,2\mu_B\right)^2\right]\nonumber\\
\label{potential}\\\nonumber
&&+\, \frac{v^2}{4}\sum_{i=1,2}\left[\left(A_i^{(1)}\right)^2 \,+\,\left(A_i^{(2)}\right)^2\,+\,\left(A_i^{(3)}\right)^2\right]\,+\,m^2\,v^2\,+\,g_4\,v^4\,+\,g_6\,v^6\,.
\eea
One consistent extremum is given by $v=0$, and all gauge fields also vanishing. This is the trivial solution. However, this solution cannot dominate the grand canonical ensemble for generic values of the chemical potential. In particular,  the scalar field theory without Chern-Simons terms ($k^{-1}\to 0$),  and at weak coupling  $(g_{4}\ll m, g_6\ll 1)$,  develops a Bose condensate when $|\mu_B| >  m$.   This non-trivial phase with $v\neq 0$ must persist when the coupling to Chern-Simons gauge fields is turned on. 
In order to arrive at a static and translationally invariant ground state, we need to find the minima of the potential energy function \eqref{potential}. We adopt a notation which is appropriate for $SU(2)$ by introducing three-vectors in the internal ``isospin" directions:
\be
{\bf A}_{\mu}\,\equiv\,\left(\langle A_\mu^{(1)}\rangle,\,\langle A_\mu^{(2)}\rangle,\,\langle A_\mu^{(3)}\rangle \right)^{ T}\qquad 
{\bf e}^{a}\,\equiv\,\left(\delta^{a,1},\,\delta^{a,2},\,\delta^{a,3}\right)^T\,.
\ee
In terms of these, the vacuum equations determining the ground state are (here the `$\times$' and `$\cdot$' symbols denote cross- and dot-products in the internal space):
\begin{eqnarray}
&& v^2 {\bf A}_y\,=\,\frac{k}{2\pi}{\bf A}_0\times{\bf A}_x\,,\qquad v^2 {\bf A}_x\,=\,\frac{k}{2\pi}{\bf A}_y\times{\bf A}_0\,,\label{eq1}\\\nonumber\\
&& -v^2\left({\bf A}_0-2\mu_B {\bf e}^3\right)\,=\,\frac{k}{2\pi}{\bf A}_x\times{\bf A}_y\,,\label{eq2}\\\nonumber\\
&&\frac{v}{2}\left[\left({\bf A}_0-2\mu_B {\bf e}^3\right)^2-\left({\bf A}_x\right)^2-\left({\bf A}_y\right)^2\right]\,=\,\frac{\partial V}{\partial v}\,.\label{eq3}
\end{eqnarray}
The  two equations in \eqref{eq1} together imply that ${\bf A}_0, {\bf A}_x$ and ${\bf A}_y$ are mutually orthogonal in the internal isospin directions, and that
\be
\left|{\bf A}_x\right|\,=\,\left|{\bf A}_y\right|\qquad \left|{\bf A}_0\right|\,=\,\frac{2\pi v^2}{|k|}\,\,,\qquad {\rm sgn}\left[\left({\bf A}_x\times {\bf A}_y\right) \cdot{\bf A}_0\right]\,=\,{\rm sgn}(k)\,.
\ee
Next, by taking a cross-product of eq.\eqref{eq2} with ${\bf A}_0$, we deduce that ${\bf A}_0\,=\, \langle A_0^{(3)}\rangle  {\bf e}^3$:
\be
\left({\bf A}_x\times{\bf A}_y\right)\times{\bf A}_0\,=\,0\, \implies\, {\bf A}_0 \times {\bf e}^3\,=\,0\,.
\ee
Finally, combining equations \eqref{eq2} and \eqref{eq3}, we obtain conditions on the magnitudes of the background field expectation values:
\bea
&&\left|{\bf A}_x\right|^2\,=\,\left|{\bf A}_y\right|^2\,=\,\frac{2\pi v^2}{|k|}\left|\langle {A}_0^{(3)}\rangle\,-\,2\mu_B \right|\\\nonumber\\
&& \left(\langle {A}_0^{(3)}\rangle\,-\,2\mu_B\right)^2\,-\,\frac{4\pi v^2}{|k|}\left|\langle{A}_0^{(3)}\rangle\,-\,2\mu_B \right|\,-\,\frac{2}{v}\frac{\partial V}{\partial v}\,=\,0\,.\label{vaceq}
\eea
To proceed further, it is useful to work with the (isospin) basis elements
\bea
&&{\bf A}_0\,=\,\eta \frac{2\pi v^2}{|k|} {\bf e}^3\,,\qquad{\bf A}_x\,=\, a_1\,{\bf e}^1 \,+ \,a_2 \,{\bf e}^2\,,\qquad
{\bf A}_y\,=\, \eta\, {\rm sgn}(k)\,\left(-a_2\,{\bf e}^1\, +\, a_1 \,{\bf e}^2\right)\,,
\nonumber
\eea
where $\eta \,=\,\pm 1$ and $a_{1,2}\in {\mathbb R}$. Using the equations of motion \eqref{eq1} and \eqref{eq2} we then find that
\be
 \eta \,=\,{\rm sgn}(\mu_B)\,, \qquad (a_1)^2\, +\, (a_2)^2\,=\,\frac{4\pi v^2}{|k|}\left(|\mu_{B}| -\frac{v^{2}\pi}{|k|}\right)\,,\qquad   |\mu_B| > \frac{\pi v^2}{|k|}\,.
 \label{E14} 
 \ee
 The classical configuration is endowed with a non-zero $U(1)_B$ charge density,
 \be
\langle j^0_B\rangle\,=\,{\rm sgn}(\mu_B) \frac{2\pi v^4}{|k|}\,,
 \ee
 with vanishing $U(1)_B$ currents.
 To calculate the scalar VEV we need the form of the tree level potential. For simplicity we set $g_6=0$. With a quartic potential there exists a unique solution\footnote{The second root for $v^2$ yields $v^2 >|\mu_B k|/2\pi$ and violates the condition in eq. \eqref{E14}.} for the vacuum expectation value \eqref{vaceq},
 \be
 v^2 = \frac{|k|}{3\pi} \left( \frac{g_4 |k|}{\pi} + {2 |\mu_{B}|} - \sqrt{ \left(\frac{g_4 |k|}{\pi} + {2 |\mu_{B}|}\right)^2\,-\,3( \mu_{B}^2 -m^2)
 } \right)\label{VEV}\,,
 \ee
 which also satisfies the condition \eqref{E14}. As expected, the VEV is real only when $\mu_B^2 > m^2$, and in the large $k$ limit when the Chern-Simons gauge fields decouple, $v^2 \simeq (\mu_B^2-m^2)/2 g_4$. This is, of course, the scalar VEV in the pure scalar theory in the Bose condensed phase. In the massless theory, the scalar VEV is controlled by the dimensionless combination $|\pi \mu_B/g_4k|$:
 \bea
 m=0:\qquad &&v^2\,=\,\frac{|\mu_B k|}{2\pi}\,f(\tilde\mu)\,,\qquad \tilde\mu\,\equiv\,\frac{\pi |\mu_B|}{g_4 |k|}\\\nonumber\\\nonumber
 && f(\tilde\mu)\,=\,\frac{2}{3} \left(\tilde\mu^{-1}+2 -\sqrt{(\tilde\mu^{-1}+2)^2-3}\right)\,,
 \eea
 where $f(\tilde \mu)$ is  monotonically increasing with $f(0)=0$ and $f(\infty)\simeq \frac{2}{3}$. A noteworthy point here is that the scalar VEV exists even when $g_4$  technically vanishes.  More generally, one may view the semiclassical limit in which the condensate is well defined as  $(g_4/\mu_B)\to 0$ and $k\to\infty$ such that $g_4 k/\mu_B$ is kept fixed.
 
\begin{figure}[h]
\begin{center}
\includegraphics[width=2in]{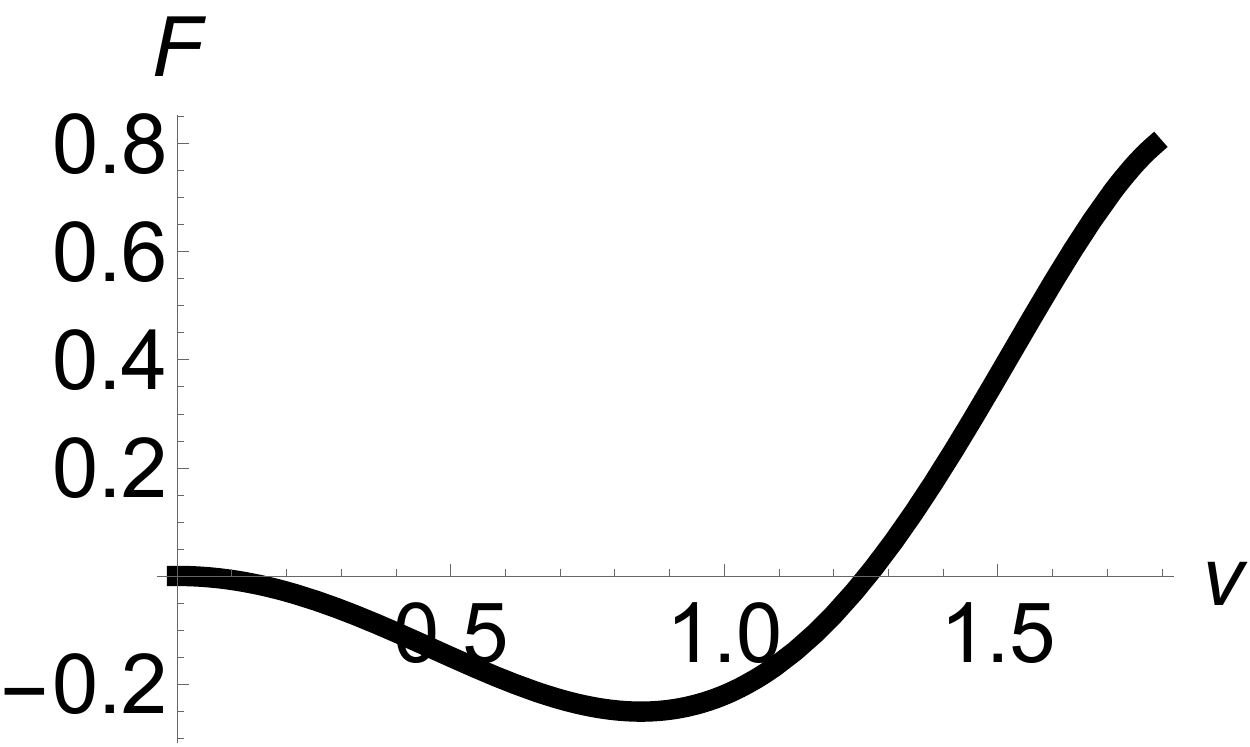}
\end{center}
\caption{ The effective potential (free energy density) as a function of the VEV $v$ for $\mu_B=1$, $m=0.5$ and $g_4=0$.
}
\label{fvsv}
\end{figure}

 \paragraph{Free energy:} For static configurations  we can compute the free energy density by evaluating the potential energy function on the ground state.   In terms of the VEV, the free energy is,
 \be
F \,=\, v^2\left[g_4v^2 +m^2 - \left(|\mu_B|-\frac{\pi v^2}{k}\right)^2\right]\,.
 \ee
 It is easy to check that (assuming $|\mu_B| > m$) the function is negative definite.
 In the massless case, the free energy of the Bose condensed phase is determined by the function $f(\tilde\mu)$:
 \be
 F\left.\right|_{m=0}\,=\,\frac{|\mu_B^3 k|}{4\pi} \frac{f(\tilde\mu)}{\tilde\mu}\left[f(\tilde\mu)-\tfrac{\tilde\mu}{2}\left(f(\tilde\mu)-2\right)^2\right]\,,
 \ee
 which is a negative definite, monotonically decreasing function of $\tilde\mu$. Therefore, in the semiclassical regime, the nontrivial vacuum dominates over the trivial one with vanishing VEVs for all fields. For instance, in the massless theory with $g_4=0$, the free energy in the Higgsed phase is
 \be
 F\left.\right|_{m=0,g_4=0}\,=\,-4\frac{|\mu_B^3 k|}{27\pi}\,,
 \ee
  valid in the semiclassical limit $k\gg 1$. Quantum corrections are parametrically suppressed in this limit. In this case the theory enters the Higgsed phase for any non-zero chemical potential, while the theory with vanishing chemical potential is conformal.   When the mass is non-zero and the chemical potential is dialed past the classical threshold value $\mu_B=m$, following a second order phase transition, the theory enters a Bose condensed Higgs phase.   The symmetric phase is unstable beyond this point. This interpretation is supported by the plot (figure \ref{fvsv}) of the free energy as a function of the VEV $v$ (taking $g_4=0$ for simplicity).
The effect of  quantum corrections at large $k$ will be to renormalize the mass  in the symmetric phase and change the threshold value of the chemical potential at which the (second order) phase transition from the symmetric to the Higgsed phase occurs. This qualitative picture may change for finite $k$ when quantum corrections are large.

 \subsection{Colour-flavour locked symmetry}
 We have found a one-parameter family of gauge field solutions parametrised by the variables $(a_1, a_2)$, satisfying a constraint \eqref{E14}. Any given realization breaks the $SU(2)$ gauge symmetry completely due to the scalar VEV which also breaks $U(1)_B$. However, the scalar VEV is left invariant by the diagonal combination of $U(1)_B$ and a  $U(1)$ subgroup of the {\em global}  $SU(2)$ colour rotations:
 \be
U(1)_B:\, \langle \Phi\rangle \,\to\,e^{i\vartheta/2} \langle \Phi\rangle \,\qquad U(1)_C:\,\Phi\,\to\,U(\vartheta)\Phi\,,\qquad U(\vartheta)\,\equiv\,e^{i\vartheta\sigma_3/2}
 \ee
While the gauge fields do not transform under $U(1)_B$, they do transform under the global $U(1)_C$ . The transformation acts on the background gauge fields $\langle A_i\rangle \,=\,\langle A_i^{(a)}\rangle t^a$ exactly as a rotation $(R)$ by a constant angle $\vartheta$ in the $x$-$y$ plane:
\bea
U(1)_C:\, \left(\begin{matrix}
\langle A_x\rangle \\ \langle A_y\rangle
\end{matrix}\right)
 \,\to\,  \left(\begin{matrix}  U(\vartheta)\langle A_x\rangle U^\dagger(\vartheta) 
\\ \\  U(\vartheta)\langle A_y\rangle U^\dagger(\vartheta)\end{matrix}\right)
 \,=\,\left(\begin{matrix} \cos \vartheta \, & \quad-\sin \vartheta
 \\\sin \vartheta\,&\quad\cos \vartheta \end{matrix}\right )
 \left(\begin{matrix} \langle A_x\rangle \\ \langle A_y\rangle \end{matrix}\right ),
\label{rotmatrix}
\eea
Therefore the vacuum gauge configuration is invariant under a global $U(1)_{B+C+R}$ symmetry which can be viewed as a  linear combination of global colour, flavour (or baryon number) and $SO(2)$ rotations in the $x$-$y$ plane.

The above observation has an important consequence. It implies that the ground state does {\em not} actually break rotational invariance\footnote{This will be corroborated by the spectrum of physical fluctuations which we extract subsequently.}, since the action of  rotations can be undone by  a gauge transformation. This is naturally reflected in the expectation values of all gauge invariant operators built from field strengths. In particular, the expectation values of single trace operators built from the chromoelectric and chromomagnetic field strengths are independent of the spatial direction or spatial component in question:
\be
\langle{\rm Tr}\left( F_{0i}\right)^{2}\rangle\,=\,-\tfrac{2\pi^3 v^6}{|k|^3}\left(\mu_B-\tfrac{v^2\pi}{|k|}\right)\,,\qquad \langle{\rm Tr}\left( F_{ij}\right)^{2}\rangle\,=\,-\tfrac{8\pi^2 v^4}{|k|^3}\left(\mu_B-\tfrac{v^2\pi}{|k|}\right)^2\,.
\ee

\section{Spectrum of fluctuations}
\label{sec3}
We now turn to the spectrum of quadratic fluctuations about the classical vacuum configuration. In the quantum theory this is reliable at weak coupling i.e. $k\gg 1$ and $\mu_B\gg g_4$. 
\subsection{The $k\to\infty$ theory} 
It is useful to first recall the situation when the Chern-Simons fields are decoupled in the limit $k\to \infty$. In this limit we have a pure scalar field theory with a global $O(4)\supset SU(2)\times U(1)_B$ symmetry. A large enough chemical potential for $U(1)_B$ leads to Bose condensation via  the scalar VEV,
\be
k\to\infty: \quad v^2 \,=\,\frac{\mu_B^2- m^2}{2 g_4}\,, 
\ee
 and the weak coupling spectrum  is readily obtained after diagonalizing the matrix of quadratic fluctuations. There are four physical excitations corresponding to the four real scalar degrees of freedom with the following dispersion relations for the frequency $\omega$ as a function of the spatial momentum ${\bf p}$, where we have set $m=0$ for simplicity:
 \bea
&& \omega^2_{{\rm I}\,(\pm)}\,=\,{\p}^2 + 3\mu^2_B\pm\mu_B\sqrt{4\p^2+9\mu^2_B}\,,\\\nonumber
\\\nonumber
 && \omega^2_{{\rm II}\,(\pm)}\,=\,{\p}^2 + 2\mu^2_B\pm 2\mu_B\sqrt{\p^2+\mu^2_B}\,.
 \eea
Two of these states are gapless\footnote{The chemical potential picks out a  $U(1)_B\simeq SO(2)\subset O(4)$ and breaks the symmetry to $SO(3)\simeq SU(2)$. The scalar condensate spontaneously breaks both the $SU(2)$ and the $U(1)_B$, and the number of Goldstone bosons is lesser than the number of broken generators, as expected  when relativistic invariance is absent \cite{nielsen-chadha, watanabe}.}. Of the two, only one has a linear dispersion relation at low momentum and corresponds to the phonon mode while the other has a quadratic dependence on the spatial momentum,
 \be
 \omega_{{\rm I}(-)}\,=\,\frac{|\p|}{\sqrt 3}+\ldots\,,\qquad  \omega_{{\rm II}(-)}\,=\,\frac{\p^2}{2\mu_B}\,+\ldots
 \ee
The presence of the second gapless mode with quadratic dependence on momentum implies that the Bose condensed ground state cannot be viewed as a 
superfluid, due to vanishing critical velocity according to the Landau criterion \cite{Landau:1941vsj, schmitt}. This picture undergoes a qualitative change  for finite large $k$. 

\subsection{Finite, large $k$}  
For any finite value of $k$, the Chern-Simons gauge fields  couple to the scalars. However, since the gauge fields are non-dynamical, the number of physical degrees of freedom remains unaltered and is given by the number of real scalars. To calculate the semiclassical spectrum we expand  in  fluctuations about the gauge and scalar VEVs,
\begin{eqnarray}
&&A_{\mu}\,=\,\langle{A}_{\mu}\rangle\,+\,{\cal A}_{\mu}\,,\qquad 
\Phi\,=\,\langle\Phi\rangle\,+\,{\delta \Phi}\,,\qquad\delta\Phi\,\equiv\,
\ \left(
\begin{matrix}
 \varphi_{1}+i \varphi_{2} \\ \,\varphi_{3}+i \varphi_{4}
\end{matrix}
\right)\,,\label{E17}
\end{eqnarray}
where $ {\cal A}_{\mu}$  and $\{\varphi_i\}$  $(i=1,\ldots 4)$ are respectively, the gauge field  and matter  fluctuations.  Substituting these into the original action \eqref{E3} and  \eqref{E4}, and expanding to quadratic order in fluctuations, 
\bea
{\cal L}^{(2)}\,=&&\delta\Phi^\dagger\,{\cal D}_\mu {\cal D}^{\mu} \,\delta\Phi\,+\, \langle\Phi^\dagger\rangle {\cal A}_\mu {\cal D}^\mu \delta\Phi - {\cal D_\mu}\delta\Phi^\dagger {\cal A}_\mu \langle\Phi\rangle +\langle\Phi^\dagger\rangle {\cal A}_\mu {\cal A}^\mu \langle\Phi\rangle\\\nonumber\\\nonumber
+&&\frac{k}{4\pi}\epsilon^{\mu\nu\lambda} {\rm Tr}\left({\cal A}_\mu {\cal D}_\nu{\cal A}_\lambda\right) \,-\,\frac{1}{2}\varphi_j\,\varphi_k \left\langle\frac{\partial^2 V}{\partial \varphi_j\partial \varphi_k}\right\rangle \,.
\eea
Here ${\cal D}_\mu$ denotes the covariant derivative with respect to the background gauge field $\langle A_\mu\rangle$:
\bea
{\cal D}_\mu \delta\Phi\,\equiv\,\partial_\mu\delta\Phi + \left(\langle A_\mu\rangle\,+\,i\mu_B\delta_{\mu,0} \right)\delta\Phi\,,\qquad {\cal D}_\mu {\cal A}_\nu\,\equiv\,\partial_\mu {\cal A}_\nu\,+\,[\langle A_\mu\rangle, {\cal A}_\nu]\,.
\eea
The main point to note here is that in the presence of the VEV for both scalars and gauge fields, all the fluctuations (matter and gauge) couple to each other at quadratic order. Due to the mixings, the physical degrees of freedom and their dispersion relations are not immediately obvious. In order to extract these, we first need to gauge-fix the action for the quadratic fluctuations.  The gauge-{\em unfixed} action would yield a degenerate matrix with vanishing determinant. In the presence of background gauge fields and symmetry breaking scalar VEVs, it is natural to adopt an $R_{\xi}$ gauge which is covariant with respect to the non-zero background gauge fields:
\bea
&&{\cal L}^{(2)}\,\to\, {\cal L}^{(2)} \,+\, {\cal L}_{\rm gf}\,,\qquad 
{\cal L}_{\rm gf}\,=\,\frac{1}{2\xi}{\rm Tr}\left({\cal D}_\mu{\cal A}^\mu\,-\xi \langle\Phi\rangle \delta\Phi^\dagger \,+\,\xi \delta\Phi \langle\Phi^\dagger\rangle  \,\right)^2\,.
\eea
The $R_{\xi}$ gauge above removes the derivative couplings between the would-be Goldstone modes and the gauge field fluctuations ${\cal A}_\mu$, and introduces a non-trivial mass matrix for them.

The determinant of the gauge-fixed fluctuation matrix then exhibits zeroes with both $\xi$-dependent and $\xi$-independent dispersion relations. The latter correspond to the physical states of the theory. In fact, these can be isolated by identifying the leading term in the large-$\xi$ expansion of the determinant of fluctuations at fixed frequency and momentum.

\subsubsection{Physical states}
We have checked numerically that the physical states inferred from the procedure above are indeed $\xi$-independent. For the $SU(2)$ theory there are precisely four physical states corresponding to the  two complex components of the scalar doublet, since the  Chern-Simons gauge fields cannot contribute any additional physical, propagating degrees of freedom. The dispersion relations for these four physical states are given by the solutions to a quartic equation in $(\omega^2, {\bf p}^2)$,
\be
\omega^8\,+\,\mu^2_BC_3\,\omega^6\,+\,\mu^4_BC_2 \,\omega^4 \,+\,\mu^6_BC_1\, \omega^2 \,+\,\mu^8_B\, C_0\,=\,0\,.\label{det}
\ee
where the $\{C_i\}$ $(i=0,\ldots3)$ are functions of dimensionless variables,
\be
C_i\,=\, C_i\left(\frac{{\bf p}^2}{\mu^2_B},\,\frac{g_4}{\mu_B},\,\frac{m^2}{\mu_B^2},\,k\right)\,,
\ee
whose explicit forms are given in \eqref{ci}. 

\paragraph{ The phonon mode:} We first recall that the $U(1)_B$ global symmetry is spontaneously broken and the corresponding Goldstone mode is the phonon. Since the remaining broken symmetries are local, the phonon should be the only massless state. This is confirmed by solving for the spectrum using \eqref{det} at ${\p=0}$ which yields\footnote{We follow the branches with the same nomenclature used for the $k=\infty$ theory. The subscripts ${\rm I}(-)$ and ${\rm II}(-)$ refer to the gapless states in that theory with linear and quadratic dispersion relations, respectively.}
\bea
{\bf p=0:}\quad&&\omega_{{\rm I}-}\,=\,0\, \qquad \qquad
\omega_{{\rm I}+}\,=\,\sqrt{m^2+6g_4v^2 -\mu_B^2 +\left(2|\mu_B|-\frac{\pi v}{|k|}\right)^2}\,,\nonumber\\\\\nonumber
&& \omega_{{\rm II}(-)}\,=\,\frac{4\pi}{|k|}\,v^2 \qquad\qquad \omega_{{\rm II}(+)}\,=\, 2|\mu_B|\,.
\eea
As expected the gapless mode with a quadratic dispersion in the $k=\infty$ theory is lifted. It is then straightforward to find the velocity of the phonon mode. In the limit of small $\omega$ and $|\p|$ we identify the coefficients of the terms quadratic in $\omega$ and $\p$ in the polynomial \eqref{det}. The resulting speed of sound  is then,
\bea
&& c_s\,=\,\left.\frac{d\omega}{d|\p|}\right|_{|\p|\to 0}\,=\,\left(1- y\right)^{1/2}\left(\frac{-15 y^2+12 y + (m^2+6 g_4 v^2)\mu_B^{-2}-1}{y^2-4 y + (m^2+6 g_4 v^2)\mu_B^{-2}-3}\right)^{1/2}\nonumber
\\\nonumber\\&& y\,\equiv\, \frac{\pi v^2}{|k\mu_B|}\,.
\eea
The scalar VEV is given in eq.\eqref{VEV}.
In the massless limit ($m=0$), the expression is purely a function of the dimensionless combination $\tilde \mu = \pi\mu_B/|g_4 k|$ introduced earlier.  In particular, the two distinct regimes of large $k$ (with $g_4$ fixed) and  small $g_4$ (with $k$ fixed), which correspond to small and large $\tilde \mu$ respectively, are distinguished by  two different limiting values for the speed of sound:
\bea
&&m=0\,,\quad \tilde \mu \ll 1\,: \qquad c_s\,=\,\frac{1}{\sqrt 3}\left(1\,+\,\frac{5\,\tilde \mu}{12}\,-\,\frac{91\,\tilde\mu^2}{96}\,+\ldots\right) \\\nonumber\\\nonumber
&& m=0 \,,\quad \tilde \mu \gg 1\,: \qquad c_s\,=\,\frac{1}{\sqrt 2}\left(1\,-\,\frac{1}{8\,\tilde\mu} \,+\,\frac{11}{128\,\tilde\mu^2}\,+\ldots\right) 
\eea
The limit of vanishing $g_4$ yields the free scalar field coupled to Chern-Simons gauge fields. In this limit the theory is conformal and therefore the speed of sound is as expected for a scale-invariant theory in $2+1$ dimensions. This is a consistency check of  the nontrivial Bose-condensed ground state we have discussed, stabilized by gauge field expectation values. It is also a consistency check on the dispersion relations for the semiclassical quadratic fluctuations.
\begin{figure}[h]
\begin{center}
\includegraphics[width=3in]{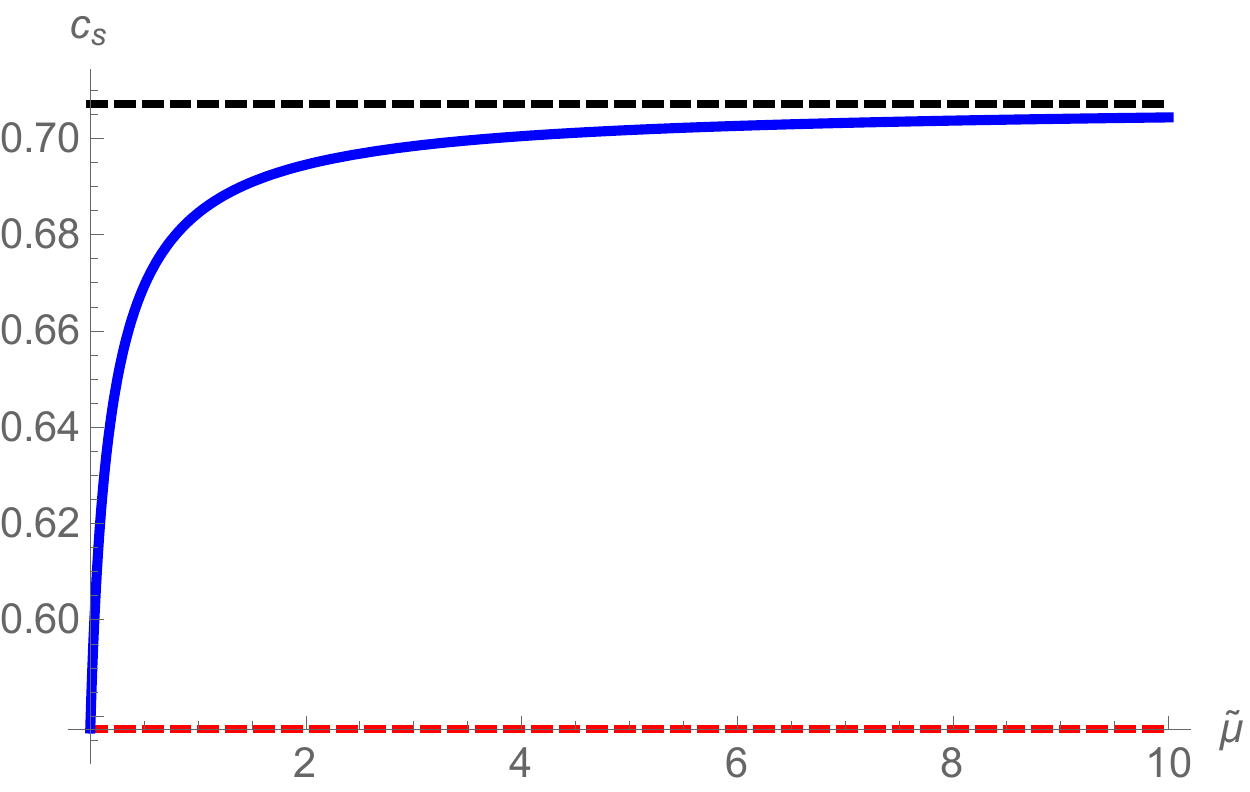}
\end{center}
\caption{ The solid blue curve shows the slope of the phonon dispersion relation at $\p =0$ as a function of $\tilde\mu = \pi|\mu_B/g_4 k|$ for the massless theory, It interpolates between $c_s=1/\sqrt{3}$ at small $\tilde\mu$ and the conformal value of $c_s = 1/\sqrt{2}$ when $g_4$ is taken to zero.
}
\end{figure}
For non-zero scalar masses the phonon velocity is a nontrivial function of both $m$ and $\mu_B$.  For instance, at large values of $k$ and all other parameters held fixed, we obtain
\bea
&&c^2_{s} \,=\,\frac{\mu_{B} ^2-m^2}{3\mu_{B} ^2-m^2}\,+\,\frac{\pi  \left(5 \mu_{B} ^2+m^2\right) (m^2 - \mu^2_B)^2 }{2 |k \mu_{B}| g_{4}   (m^2 -\mu^2_B)^2 }+ {O}(1/k^2)\,.
\eea
The expression can be rewritten as a function of the two dimensionless parameters $\tilde \mu\,=\,\pi \mu_B/g_4 |k|$ and 
$\tilde m\equiv \pi m/g_4 |k|$.
\paragraph{Level crossing:} The perturbative spectrum in the regime of small $\omega$ and $\p$ displays an interesting feature. This is a nontrivial consequence of crossing of energy levels which occurs as we tune the Chern-Simons level from $k=\infty$ to finite (large) values. This unavoidable crossing is between the phonon ($\omega_{\rm I (-)}$  branch) and the light state with energy $\omega_{\rm II(-)}$ which happens to be gapless with quadratic dispersion relation at $k=\infty$, but acquires a small gap $\sim 4\pi v^2/k$ at large $k$.  The crossing is accompanied by off-diagonal mixings between these two fluctuations. In the low energy, long wavelength limit $\omega, |\p| \ll \mu_B$ (where we are ignoring  $m$ for simplicity) it should suffice to focus attention on the two-level system comprising of the two lightest excitations. In this limit, the gapped modes only yield an overall multiplicative constant in the fluctuation determinant which takes the approximate form,
\be
\left(\omega^2 - c_s^2 \p^2\right)\left(\omega^2-\frac{\p^4}{4\mu^2}\,-\,\delta\right)\,-\,\varepsilon\, \p^4\,=0\,.\label{mixing}
\ee
The mixing term $\varepsilon\sim k^{-1}$, whilst  the gap generated for the branch $\omega_{\rm II(-)}$ with quadratic dispersion scales as  $\delta\sim k^{-2}$,  both  vanishing in the large $k$ limit.  The mixing must necessarily be momentum dependent so that the gapless phonon mode persists as a Goldstone boson for the broken $U(1)_B$. At low momentum the leading such contribution scales as $\p^4$ (using eq.\eqref{ci}). 
\begin{figure}[h]
\begin{center}
\includegraphics[width=2.3in]{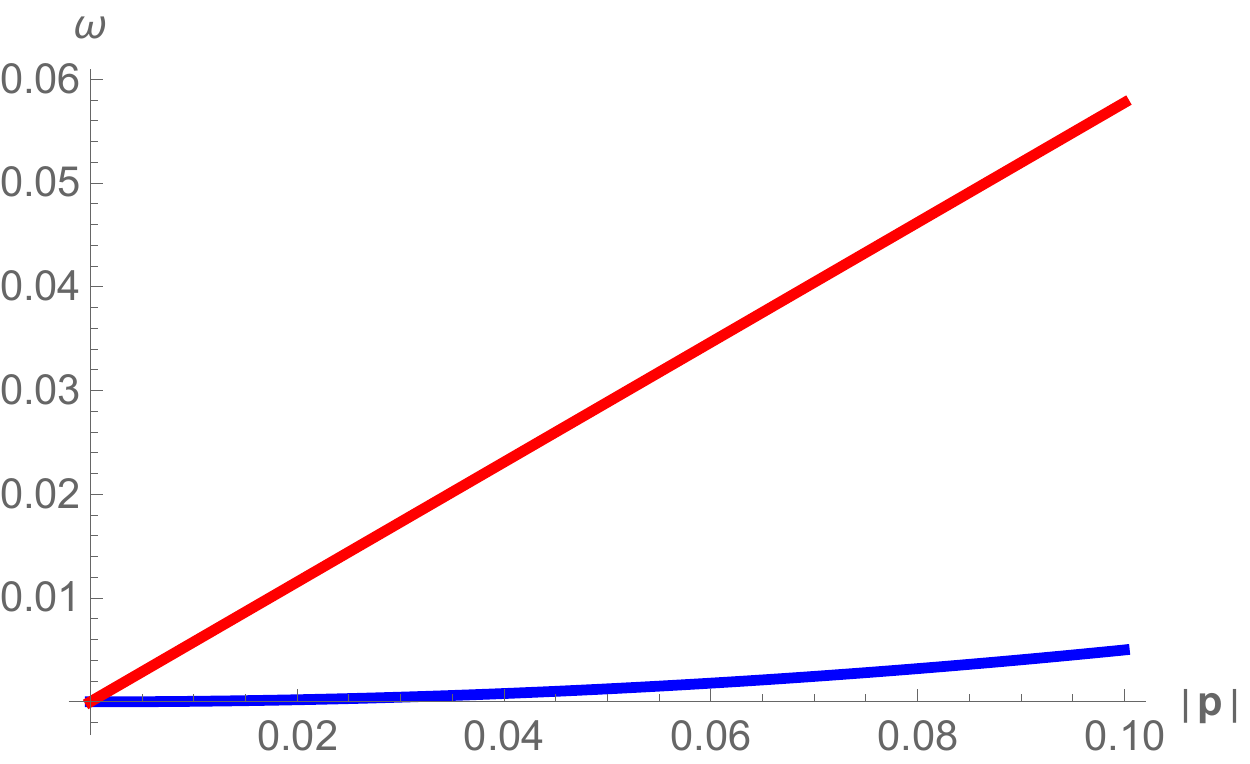}\hspace{0.8in}\includegraphics[width=2.3in]{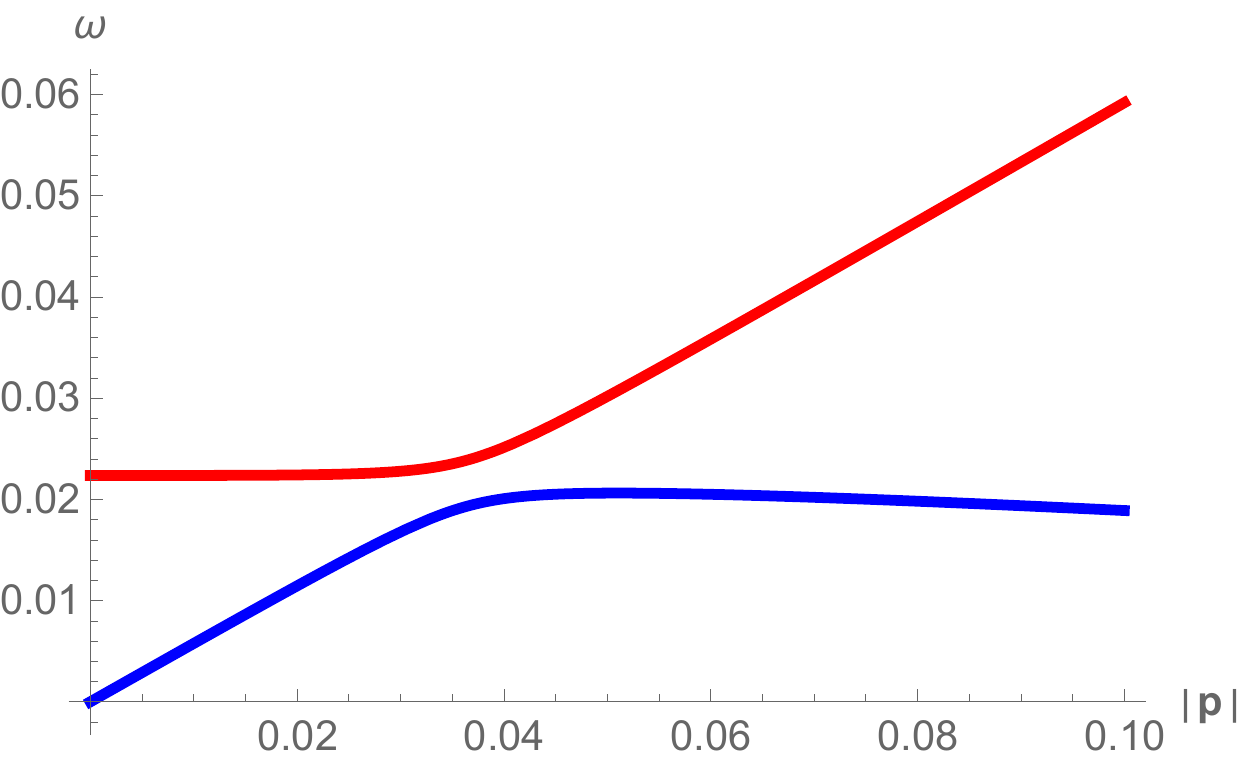}
\end{center}
\caption{ {\bf Left:} The $k=\infty$ theory displays two gapless excitations with linear and quadratic dispersion relation. {\bf Right:} At finite large $k$, the two modes potentially cross, and also mix. The diagonalized modes display  a splitting and ``repulsion" resulting in a local maximum in the phonon branch.
}\label{cross}
\end{figure}
The new solutions to \eqref{mixing} provide a qualitative description of the perturbed light spectrum at large, finite $k$.  
In particular, as shown in figure \ref{cross}, the two branches do not cross and the phonon branch displays a ``maxon" or a local maximum in its dispersion relation.  
For non-zero $\varepsilon$ the two dispersion relations (viewed as functions of $\p^2$) have a branch-point in the complex plane. For small enough $\varepsilon$, the location of the maximum in $\omega_{\rm I(-)}$ is close to the putative interesection point of the two curves.
The presence of this local maximum implies the existence of a ``roton" minimum since all dispersion curves must eventually increase linearly at large $|\p|$ consistent with UV relativistic invariance. 

\subsection{Roton minimum and complete spectrum}
\begin{figure}[h]
\begin{center}
\includegraphics[width=2.3in]{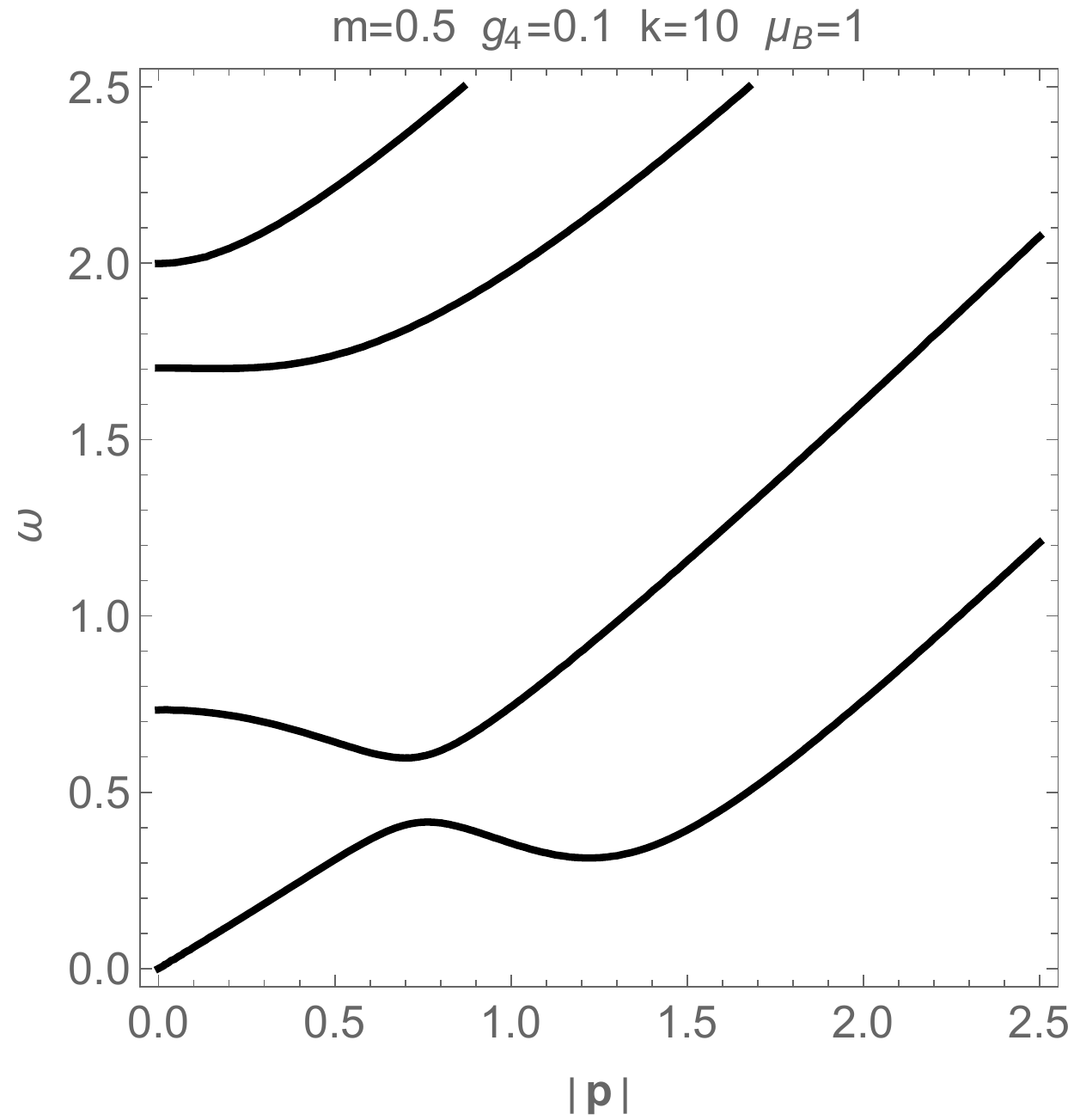}\hspace{0.3in}\includegraphics[height=2.3in]{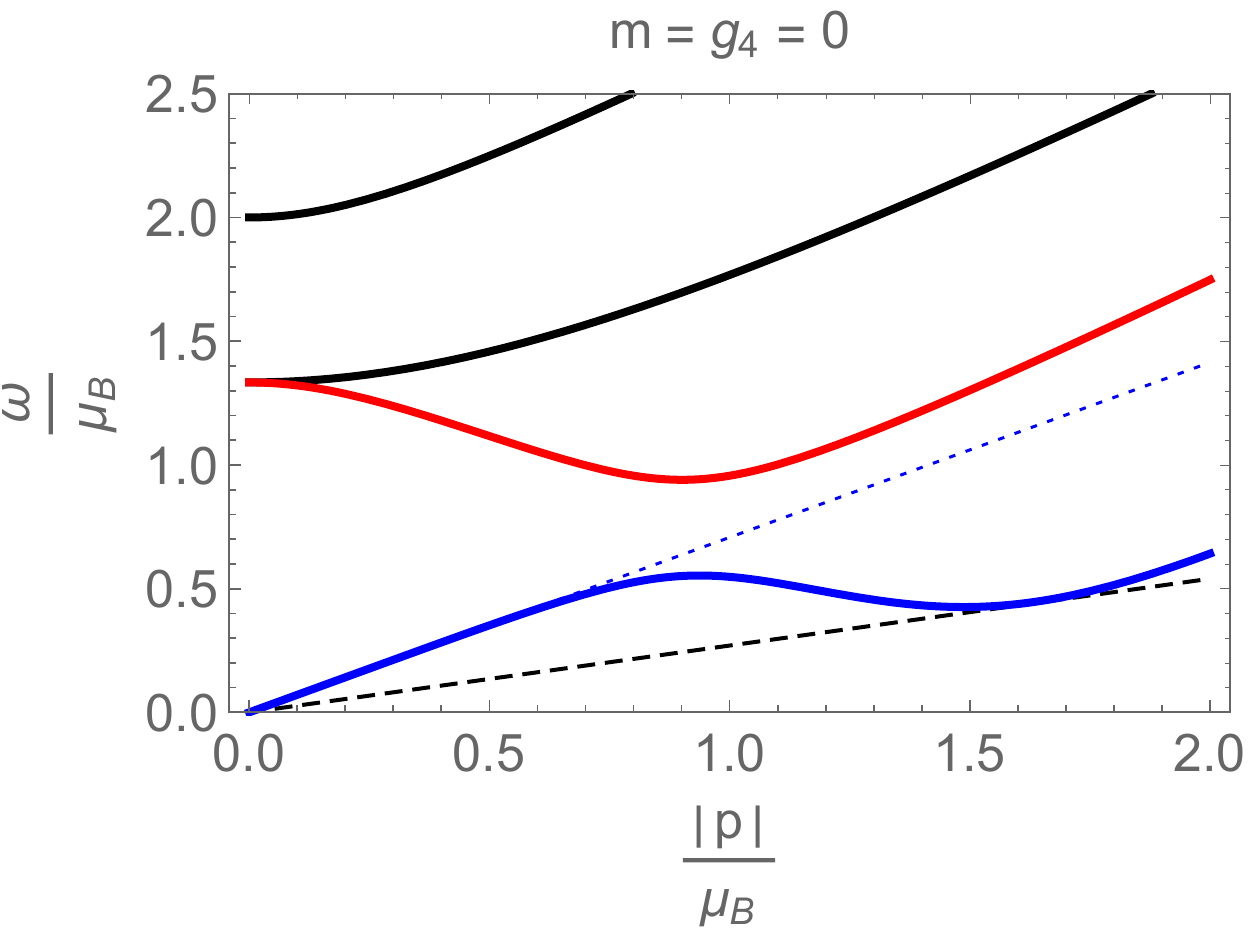}
\end{center}
\caption{ \small{{\bf Left}: The physical spectrum for the massive case. {\bf Right:} In the scale invariant situation, the dispersion relations have a universal form, independent of $k$ and $\mu$. The dotted blue line with slope $1/\sqrt 2$ matches the phonon velocity at low momentum.
}}
\label{spectmassive}
\end{figure}

Our main observation is that for any (large) finite value of $k$, consistent with being in the semiclassical regime the phonon branch always displays a roton minimum. 
At large $k$ and fixed $g_4$, the position of the maximum can be estimated quite easily. It sits close to the potential intersection point of the dispersion curves for $\omega_{\rm II(-)}$ and $\omega_{\rm I (-)}$. In the large $k$ regime, the former is flat, $\omega_{\rm II(-)}\,\approx\, 4\pi v^2/|k|$, while the latter is linear, $\omega_{\rm I(-)} \,\approx\, |\p|/\sqrt{3}$, and their putative intersection is at 
\be
k\gg 1,\,g_4\,{\rm fixed}: \qquad \left(\rm \omega_{\rm max},\, |\p|_{\rm max}\right)\,\approx\,\left(\frac{4\pi v^2}{|k|}, \, \frac{4\pi v^2\sqrt{3}}{|k|}\right)\,.
\ee
\begin{figure}[h]
\begin{center}
\includegraphics[width=2.8in]{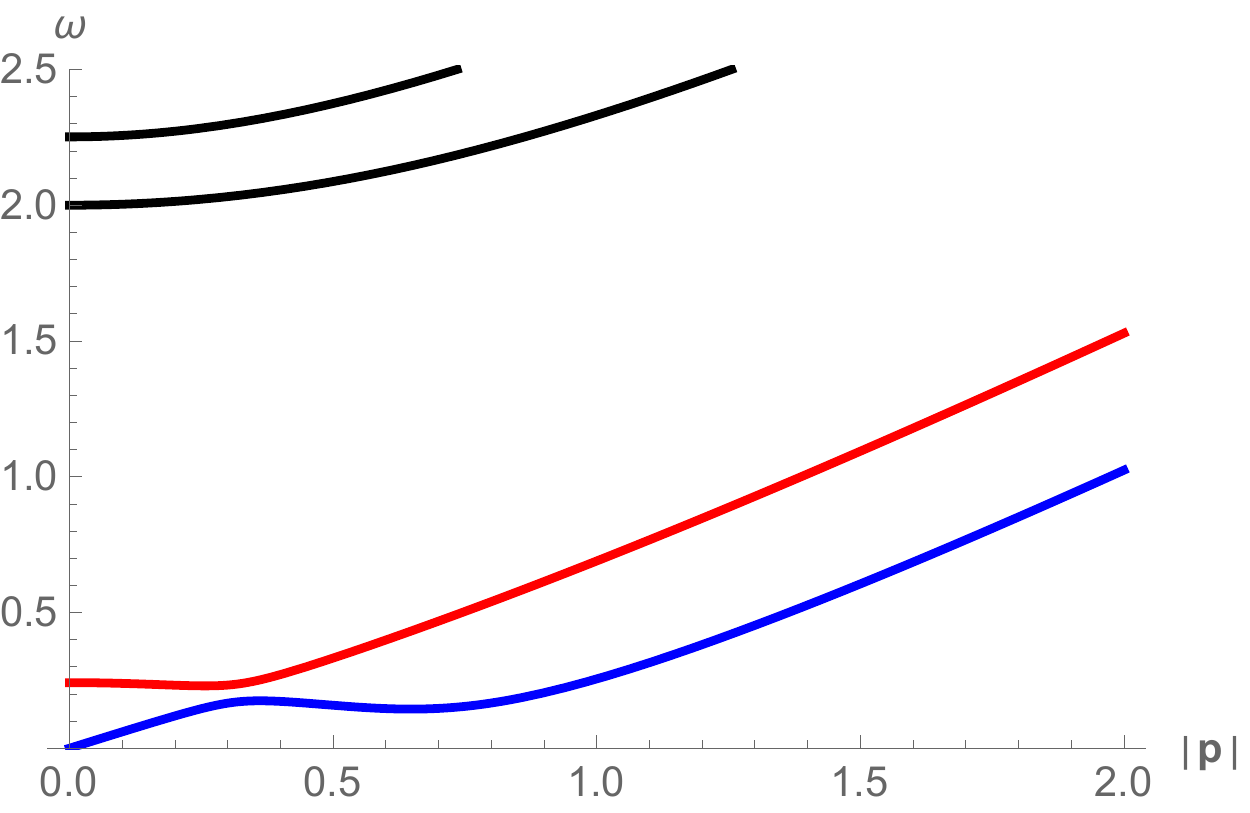}\hspace{0.1in}\includegraphics[width=3.0in]{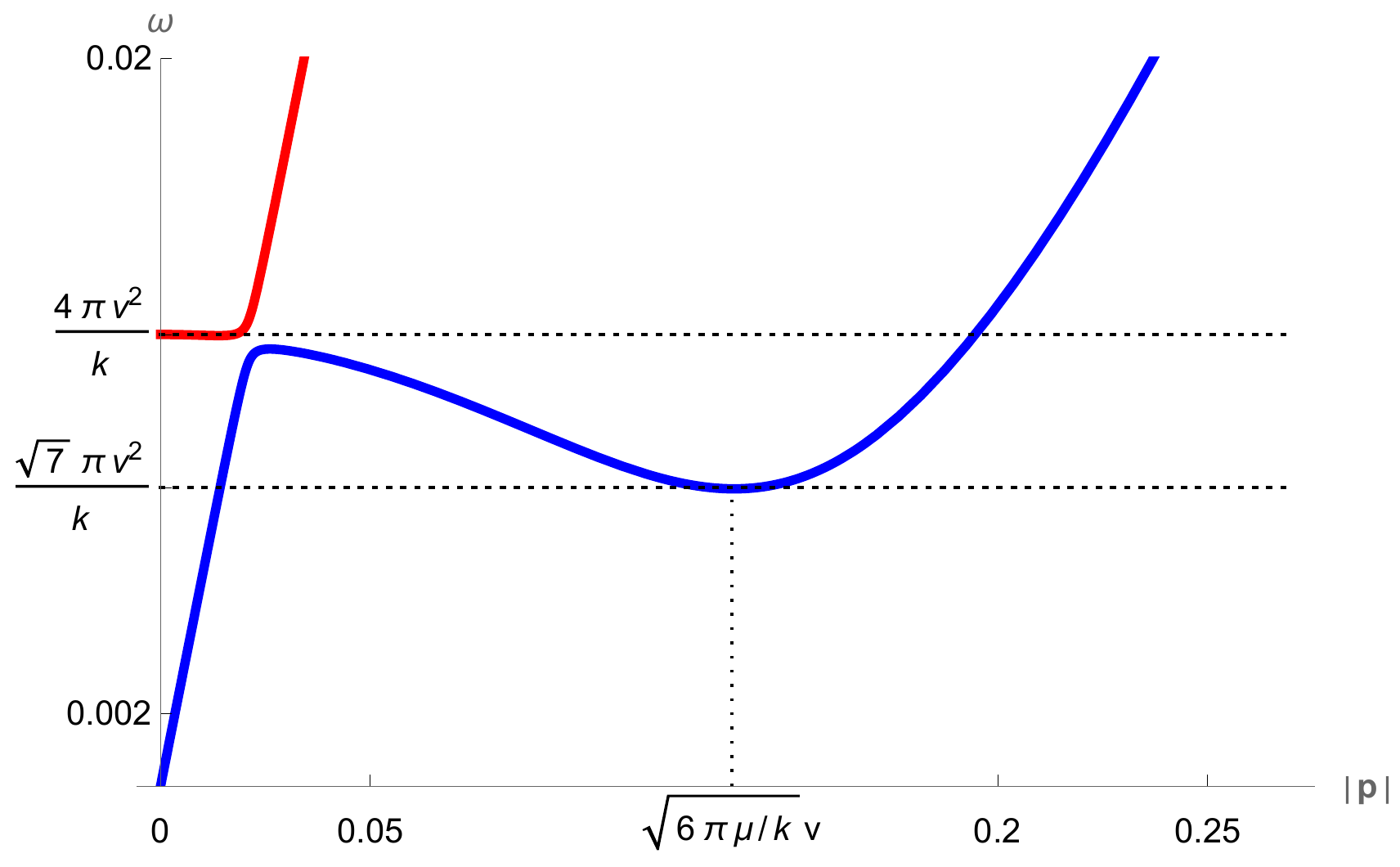}
\end{center}
\caption{ \small{{\bf Left:} The spectrum for $m=0$ with $\tilde\mu =\pi \mu_{B}/g_4 k =\pi/20 $. {\bf Right:} The two lightest states, including the phonon-roton branch (blue) with $\tilde\mu=\pi/500$. The dotted lines indicate the large-$k$ limiting values of the roton maximum and minimum.
}}
\label{largek}
\end{figure}
On the other hand, the location of the roton minimum is more subtle. In the large $k$ theory we expect the minimum to be located at parametrically small values close to the origin.  In fact, it turns out that $\omega_{\rm rot} \sim k^{-1}$ whilst  $\p_{\rm rot} \sim k^{-1/2}$. This can be checked by first performing the scaling 
\be
\omega\,=\,\frac{1}{k}\, \varpi\qquad\qquad \p\,=\, \frac{1}{\sqrt k}\,\varrho\,,
\ee
then substituting into the fluctuation determinant \eqref{ci}, and  the expression for $\omega'(\p)$  by differentiating \eqref{ci}. Subsequently, setting the determinant and $\omega'(\p)$ to zero, and then taking the large $k$ limit, we find (setting $m=0$ for simplicity):
\bea
&&3 \varrho^4\,-\,24 \pi  \mu_B  \varrho^2 v^2\,+\,4 \mu_B^2 \left(16 \pi ^2 v^4-\varpi ^2\right)\,=\,0\\\nonumber\\\nonumber
&&\varrho^4\,-\,12 \pi  \mu_B  \varrho^2 v^2\,+\,4 \mu_B^2 \left(16 \pi ^2 v^4-\varpi ^2\right)\,=\,0\,.
\eea
The solutions to these yield the roton minimum at large $k$ for the massless theory:
\be
k\gg 1: \qquad \left(\omega_{\rm rot},\,|\p|_{\rm rot}\right)\,=\,\left(\frac{\sqrt 7 \pi v^2}{k},\,\sqrt{\frac{6\pi \mu}{k}} \,v\right)\,,
\ee
where the VEV is given by \eqref{VEV} with $m=0$. The results for the roton minimum and maximum
agree perfectly with the numerical curves for the phonon-roton branch at large $k$, displayed in figure \ref{largek}. The qualitative nature of the dispersion relations persists for all values of $m$, $\mu_B$ and $g_4 k$. Figure \ref{spectmassive} shows the relevant plots for one non-zero value of $m$.
\paragraph{Critical case with $g_4=m=0$:} A nontrivial aspect of the Bose condensed ground state is that all generic features of the spectrum of fluctuations persist even when $g_4=m=0$ (and $\mu_B\neq 0$) so that the classical theory is scale invariant.  The determinant of physical fluctuations \eqref{ci} simplifies greatly, and the relevant dispersion relations are obtained from its zeroes:
\bea
&&\tilde p\,\equiv\,\frac{|\p|}{\mu_B}\,\qquad \tilde \omega\,\equiv\,\frac{\omega}{\mu_B}\,,
\label{critdisp}\\\nonumber\\\nonumber
&& \tilde p^8\,-\,\tilde p^6 \left(4\tilde\omega ^2\,+\,\tfrac{28}{9}\right)\,+\,\tilde p^4 \left(  6\tilde \omega^4\,-\,\tfrac{4}{3}\tilde \omega ^2\,+\,\tfrac{160}{81}\right)\,-\, \tilde p^2 \left(4\tilde\omega ^6\,-\,12 \tilde \omega ^4+\tfrac{992}{81}\tilde\omega ^2-\tfrac{512}{81}\right)\\\nonumber\\\nonumber&&+\,\tilde \omega ^8\,-\,\tfrac{68 }{9}\tilde\omega ^6\,+\,\tfrac{1408 }{81}\tilde\omega ^4\,-\,\tfrac{1024 }{81}\tilde\omega ^2\,=\,0\,.
\eea
At zero momentum the energies of the four physical states are:
\be
\omega_{\rm I (-)}=0\,,\qquad\omega_{\rm I (+)}\,=\,\frac{4\mu_B}{3}\,,\qquad\omega_{\rm II (-)}\,=\,\frac{4\mu_B}{3}\,,\qquad \omega_{\rm II (+)}\,=\,2\mu_B\,,
\ee
so that two of the massive states become degenerate, 
whilst the roton maximum and minimum are at
\be
\left(\omega_{\rm max},\,|\p|_{\rm max}\right)\,=\,\left(0.553\mu_B,\,0.937\mu_B\right)\,,\qquad\left(\omega_{\rm rot},\,|\p|_{\rm rot}\right)\,=\,\left( 0.426\mu_B,\,1.487 \mu_B\right)\nonumber
\ee
We expect these results to be stable against quantum corrections for large enough $k$, which is the only small parameter in the system. It is interesting and somewhat unexpected (given that the roton minimum is often attributed to the presence of  a new scale)  that the roton persists in the theory where the chemical potential is the only dimensionful scale. 
\subsection{Landau critical velocity}
According to Landau's criterion, for a nonrelativistic superfluid flowing with velocity $v_s$ (with respect to a vessel or capillary), when the velocity exceeds a critical value \cite{schmitt} given by 
\be
v_{\rm crit} \,=\, {\rm min}_{|\bf p|} \left(\frac{\omega(\p)}{|\p|}\right)\,\implies \frac{\partial \omega}{\partial |\p|}\,=\,\frac{\omega}{|\p|}\,,
\ee
the fluid loses energy through dissipation and the superfluid phase can be wholly or partially destroyed e.g. by a condensate of rotons \cite{pitaevskii84, voskresenskii93}. In particular, \cite{pitaevskii84} argues for the appearance, within superfluid ${}^4{\rm He}$ flows, of a one dimensional periodic structure at rest with  respect to the walls so that the superfluidity criterion is not violated. The Landau criterion is derived by boosting the Bose condensate in the ground state along a particular direction (say the $+x$-axis) with a velocity $v_s$, and considering excitations that could reduce or dissipate the energy of the moving condensate. In the frame where the condensate has velocity $v_s$, the energy of a backscattered nonrelativistic excitation with momentum $\p$, causing dissipation  from the condensate, must  satisfy
\be
\omega(|\p|)\,-\,v_s|\p| <0,
\ee
where the second term is the result of  transformation under  the Galilean boost.
The critical value of the superfluid velocity is then given as $v_{\rm crit}\,=\,{\rm min}(\omega/|\p|)$. The arguments can also be carried out in the appropriate relativistic context (e.g.\cite{voskresenskii93, schmitt}). The critical velocity is inferred from the slope of the straight line passing through the origin and tangent to the dispersion curve for the phonon-roton branch (see dashed black line in figure \ref{spectmassive}).

\begin{figure}[h]
\begin{center}
\includegraphics[height=2.0in]{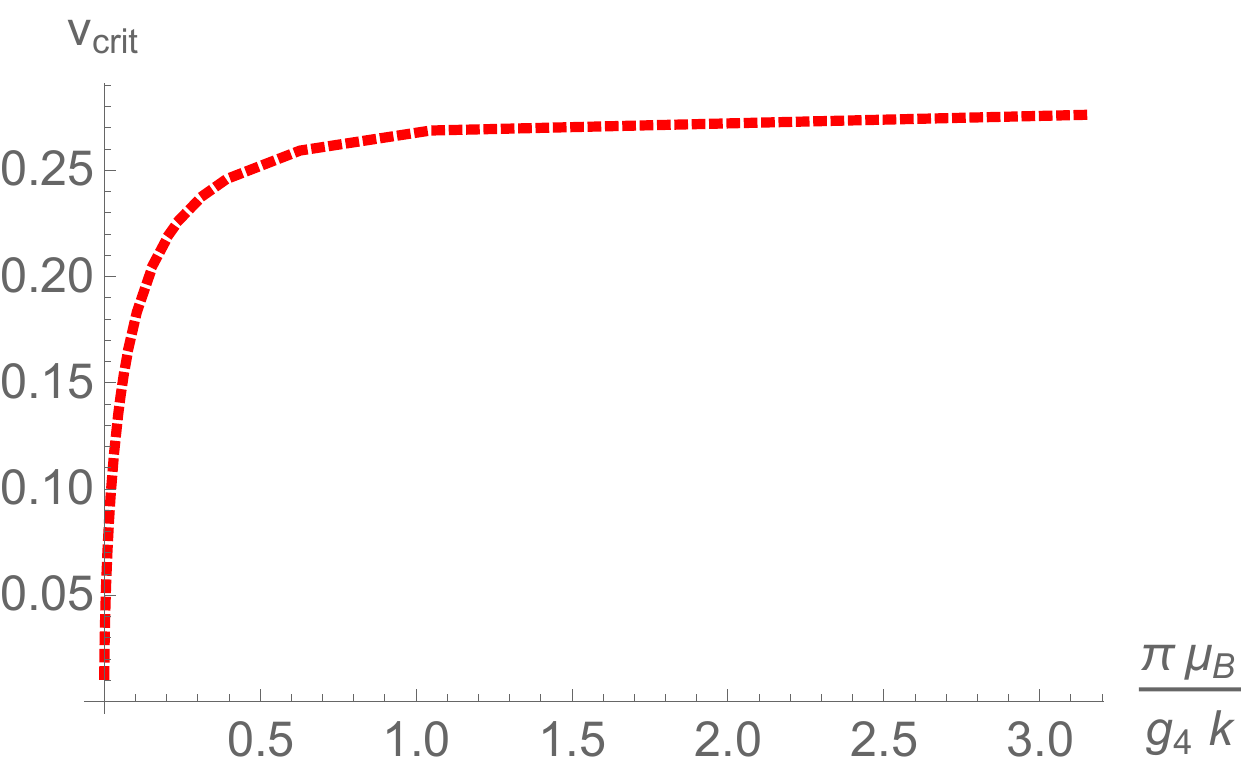}
\end{center}
\caption{ \small{The Landau critical velocity as a function of the dimensionless parameter $\pi\mu_B/g_4 k$ in the theory with $m=0$.}}
\label{vcrit}
\end{figure}

 The behaviour of the critical velocity as a function of $\mu_B/g_4 k$ in the massless theory is shown in figure \ref{vcrit}.
 At large $k$, the critical velocity vanishes as $1/\sqrt{k}$, and approaches a constant value, $v_{\rm crit}\approx 0.27$, in the theory with $g_4=0$.
 \subsection{The $U(2)_k$ theory}
\begin{figure}[h]
\begin{center}
\includegraphics[height=2.0in]{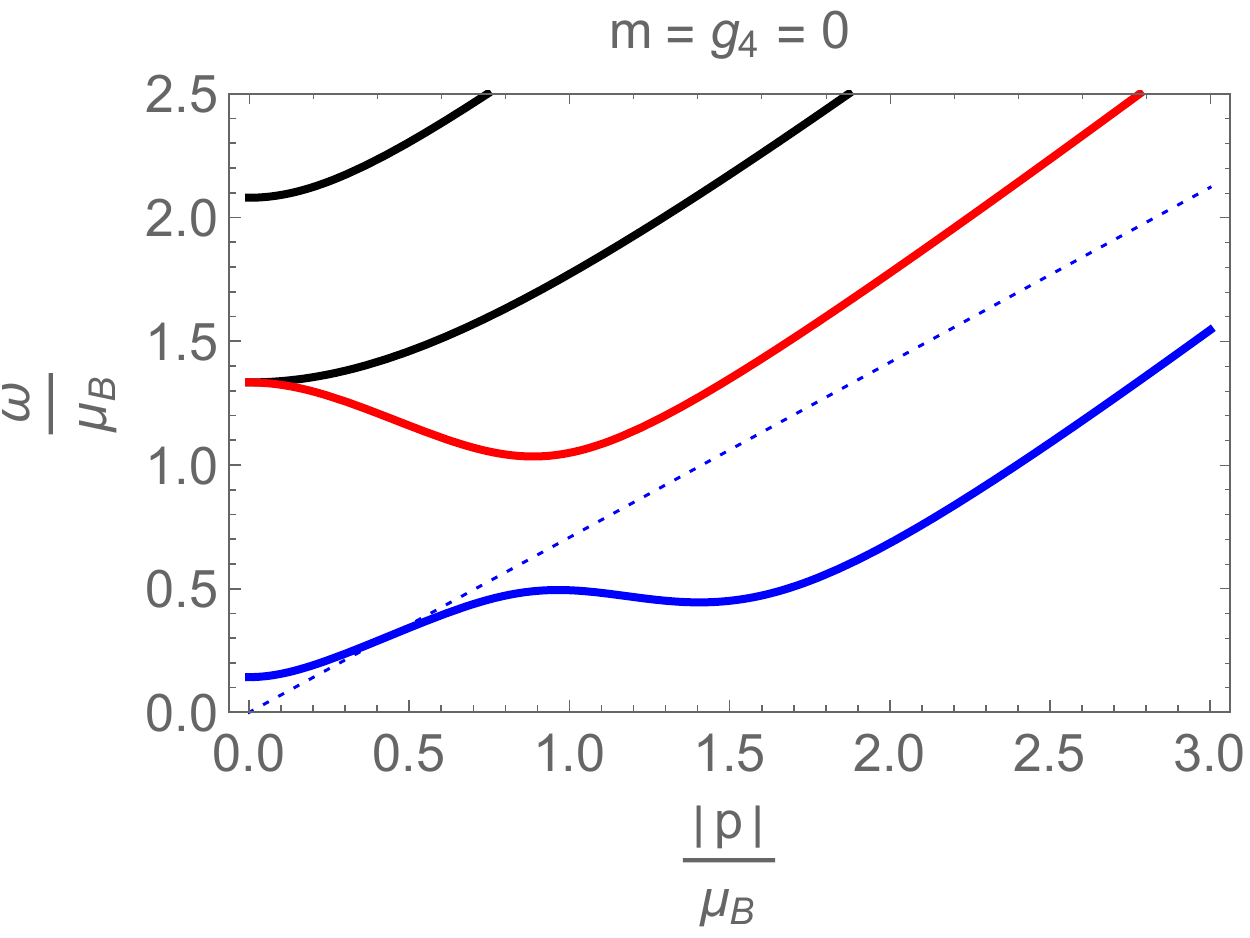}
\end{center}
\caption{ \small{The semiclassical spectrum of the $U(2)\simeq SU(2)\times U(1)$ theory. All states are gapped. Nevertheless, the lightest state displays a roton-like minimum. The dotted blue line passing through the origin with slope $1/\sqrt 2$ is shown to emphasize the absence of  phonon-like linear dispersion.}}
\label{u2}
\end{figure}

 It is interesting to note the qualitative difference between $SU(2)$ and $U(2)$ gauge groups. In the latter case the $U(1)_B$ symmetry is gauged and the chemical potential is synonymous with a fixed background expectation value for the temporal component of the abelian gauge field.   The classical vacuum equations are satisfied by the same configuration as in the $SU(2)$ theory. The condensates of the scalar and gauge fields break both the $SU(2)$ and $U(1)_B$ local symmetries to a diagonal $U(1)$. 
 Since all symmetries are local we expect only massive physical states.   We obtain the physical fluctuations by employing Coulomb gauge for the abelian gauge field, and retaining the covariant $R_\xi$ gauge-fixing for the $SU(2)$ part. The situation with $m=g_4=0$ suffices to demonstrate the existence of the gap. In this case, the dispersion relations of the four physical states can be obtained from the roots of the following polynomial in  $(\tilde\omega,\,\tilde p)\,=\,\left(\omega/\mu_B, \,|\p|/\mu_B\right)$:
 \bea
 &&\tilde p\,\equiv\,\frac{|\p|}{\mu_B}\,\qquad \tilde \omega\,\equiv\,\frac{\omega}{\mu_B}\,,\\\nonumber\\\nonumber
&& \tilde p^8\,-\,\tilde p^6 \left(4\tilde\omega ^2\,+\,\tfrac{224}{81}\right)\,+\,\tilde p^4 \left(  6\tilde \omega^4\,-\,\tfrac{64}{27}\tilde \omega ^2\,+\,\tfrac{512}{243}\right)\,-\, \tilde p^2 \left(4\tilde\omega ^6\,-\,\tfrac{352}{27} \tilde \omega ^4+\tfrac{3328}{243}\tilde\omega ^2-\tfrac{4096}{729}\right)\\\nonumber\\\nonumber&&+\,\tilde \omega ^8\,-\,\tfrac{640 }{81}\tilde\omega ^6\,+\,\tfrac{4544 }{243}\tilde\omega ^4\,-\,\tfrac{10240 }{729}\tilde\omega ^2\,+\,\tfrac{16384}{59049}.
 \eea
 Unlike the $SU(2)$ theory \eqref{critdisp} we see that $\tilde\omega\,=\,\tilde p \,=\,0\ $ is no longer a solution. All states are gapped at $\p =0$, with the energies given by $\tilde \omega^2 = 16/9, 16/9, (22\pm5\sqrt{19})8/81$. The dispersion relations for non-zero $\p$ are shown in figure \ref{u2}.

 \section{The $SU(N>2)$ case}
 \label{sec4}
 We now generalize the above analysis for Chern-Simons scalar theory with $SU(N) $ gauge group. 
 We use lower case subscripts and superscripts, $(p,q,r\ldots)$ to label  fundamental and  antifundamental representation indices. The gauge covariant derivative is defined to include the chemical potential as a timelike background gauge field:
  \begin{eqnarray}
(D_{\mu})_{p}^{\ q}\,=\,\delta_{p}^{\ q}\,\partial_{\mu}\,+\,(A_{\mu})_{p}^{\ q}\,+\,i\mu_{B}\, \delta_{p}^{\ q}\,\delta_{\mu}^{0}\,. 
\end{eqnarray}
For general $N$, it is useful to define the quartic coupling so that a consistent large $N$ limit can be taken if necessary.  The potential contributions involving both gauge and scalar fields can be put together so that,
\begin{eqnarray}
V_{\rm CS}\,+\, V_{\rm scalar}&&=\, -\,\frac{k}{4\pi}\,\frac{2}{3}\,{\rm Tr}\left(A_\mu A_\nu A_\rho\right)\epsilon^{\mu\nu\rho}\,-\,
\Phi^{\dagger}\left(A^\mu + i\mu_B\eta^{\mu 0}\right)\left(A_\mu + i\mu_B\delta_{\mu}^0\right)\Phi\,\nonumber\\\label{sunpot}\\\nonumber
&&
+\,m^{2}\Phi^{\dagger}
\Phi\,+\,\frac{g_{4}}{N}(\Phi^{\dagger}\Phi)^{2}\,.
\end{eqnarray}
Assuming that the scalar obtains a vacuum expectation value, we can always use $SU(N)$ gauge rotations to place the VEV in the $N$-th component,
\begin{eqnarray}
\langle\Phi_{p}\rangle\,=\,\sqrt{N}\,v \,\delta_{p,N}\label{sclarvev}\,.
\end{eqnarray}
We have scaled out a factor of $\sqrt N$ in anticipation of the expected scaling in the  large-$N$ limit of vector models. In particular, the action for the matter fields should be ${\cal O}(N)$ in the large-$N$ limit.
The choice of scalar VEV leaves a residual $SU(N-1)$ gauge symmetry, which is then completely broken by the gauge field backgrounds in the ground state. In order to obtain the correct matrix equations of motion, we vary the action \eqref{sunpot} subject to a tracelessness condition for $SU(N)$ gauge fields, implemented by Lagrange multipliers $\Lambda^{0,1,2}$:
\be
V_{\rm CS}\,+\,V_{\rm scalar}\,\to\,V_{\rm CS}\,+\,V_{\rm scalar}\,+\,\Lambda^\mu {\rm Tr} (A_\mu)\,.
\ee
\subsection{Vacuum configuration}
The complete vacuum equations extremizing the potential function are:
\bea
&&-\frac{k}{4\pi}\, [A_\mu,A_\nu]\epsilon^{\mu\nu\lambda}\,-\,\left\{\Phi\Phi^\dagger,\,
\left(A^\lambda\,+\,i\mu_B\,\eta^{\lambda 0}{\bf 1}\right)\right\}\,+\,\Lambda^\lambda\,{\bf 1}\,=\,0\,,\qquad {\rm Tr} A_\mu\,=\,0\,,\nonumber\\\\\nonumber
&& -\,
(A_{\mu})_{N}^{p}(A^{\mu})_{p}^{N}\,+\,2i\mu_{B}(A_0)_{N}^{N}\,+\,(m^{2}-\mu_{B}^{2})\,+\,2g_{4}v^2\,=\,0\,.
\eea
The matrix $\Phi\Phi^\dagger$ is a projector, and given that the scalar VEV can be rotated into the lowest component, it has only one non-zero element,
\be
(\Phi\Phi^\dagger)_p^{\ q}\,=\,N\,\delta_{p,N}\,\delta^{q,N}\,v^2\,.
\ee
The Lagrange multipliers $\{\Lambda^\lambda\}$ are determined by taking the trace of each of the respective equations of motion so that, 
\be
\Lambda^\lambda\,=\,2v^2 \left[\left(A^\lambda\right)_N^{\ N}\,+\,i\mu_B\,\eta^{\lambda 0} \right]\,.
\ee
Following the hint provided by the $SU(2)$ vacuum solution, we take $A_0$ to be diagonal (note that  residual $SU(N-1)$ rotations can be used to diagonalize an $(N-1)\times (N-1)$ block of $A_0$). It then follows that the commutator $[A_x, A_y]$ must be diagonal $\sim {\rm diag}(1,1,\ldots, 1-N)$.  
This is reminiscent of the $N$-dimensional  representation of the $SU(2)$ algebra, where the off-diagonal ladder operators commute to yield a diagonal matrix. Motivated by this similarity, we find a simple solution for the Chern-Simons equations of motion:
\bea
&&\langle A_x\rangle_1^{\ q}\,=\, i\alpha\, \delta^{q,2}\,,\qquad \langle A_x\rangle_N^{\ q}\,=\,i\alpha\sqrt{N-1}\, \delta^{q,N-1}\\\nonumber\\\nonumber
&&\langle A_y\rangle_1^{\ q}\,=\, \alpha\, \delta^{q,2}\,,\qquad \langle A_y\rangle_N^{\ q}\,=\, -\alpha\sqrt{N-1}\, \delta^{q,N-1}\\\nonumber\\\nonumber
&&\langle A_x\rangle_p^{\ q}\,=\,i\alpha \left(\sqrt p\,\delta^{q, p+1}\,+\,\sqrt{p-1}\,\delta^{q,p-1}\right)\,,\qquad p\,=\,2,\ldots N-1\,\\\nonumber\\\nonumber
&&\langle A_y\rangle _p^{\ q}\,=\,\alpha \left(\sqrt p\,\delta^{q, p+1}\,-\,\sqrt{p-1}\,\delta^{q,p-1}\right)\,,\qquad p\,=\,2,\ldots N-1\,\\\nonumber\\\nonumber
&& \langle A_0\rangle_p^{\ q}\,=\, i\beta \left(\frac{1}{N}\delta_{p}^{\ q}\,-\,\delta_{p,N}\delta^{q,N}\right)\,, \qquad p,q\,=\,1\ldots N\,,
\eea
where the constants $\alpha$ and $\beta$ are determined by the VEV and chemical potential as,
\be
\alpha\,=\,\frac{\beta}{\sqrt N}\sqrt{\frac{\mu_B}{\beta}\,- \,\frac{N-1}{N}}\,,\qquad \beta\, =\, \frac{v^2}{\kappa}\,,\qquad \kappa\,\equiv\, \frac{k}{2\pi N}\,.\label{alphabeta}
\ee
The equation of motion for the scalar VEV (discarding the trivial extremum)  is then given by,
\be
-\frac{3}{\kappa^2}\left(1\,-\,\tfrac{1}{N}\right)^2\,v^4\,+\,v^2\left[2g_4\,+\,\frac{4\mu_B}{\kappa}\left(1-\tfrac{1}{N}\right)\right]\,-\,\left(\mu_B^2\,-\,m^2\right)\,=\,0\,.
\ee
Solving as a quadratic in $v^2$, only one solution is physical\footnote{The second root  yields $v^2 > \kappa\mu_B N/(N-1)$ which would render $\alpha$ imaginary. In addition, this solution does not have a smooth $k\to \infty$  limit.} and matches smoothly onto the semiclassical ($\kappa\gg 1$) limit:
\be
v^2\,=\,\tfrac{N\kappa}{3(N-1)}\left[\tfrac{g_4N\kappa}{(N-1)}\,+\,2\mu_B\,-\,\sqrt{\left(\tfrac{g_4N\kappa}{N-1}\,+\,2\mu_B\right)^2\,-\,3(\mu_B^2-m^2)}\right]\,.
\ee
This agrees precisely with the result \eqref{VEV}  for $N=2$ after we perform the rescalings, $v\to v/\sqrt{N}$ and $g_4 \to g_4 N$,  required to match the conventions adopted in our analysis of the $SU(2)$ theory. It is also worth remarking that the $N\to\infty$ limit, keeping $\kappa$ and $g_4$ fixed, can be readily taken and $v$  remains finite in this limit.

For the free massless scalar coupled to Chern-Simons fields ($m=g_4=0$), we obtain 
\be
v^2\,=\,\kappa \frac{N \mu_B}{3(N-1)}\,,\qquad \alpha\,=\,\frac{\mu_B}{3}\sqrt{\frac{2}{N-1}}\,.
\ee

\subsection{Interpretation as quantum Hall droplet state}
The vacuum configuration breaks the $SU(N)$ gauge symmetry completely. The scalar field VEV also breaks the global $U(1)_B$ spontaneously and therefore the spectrum must yield a massless phonon mode. As seen previously in the $SU(2)$ theory, the classical background is left invariant by a diagonal combination of $U(1)_B$, global colour and spatial rotations.  An $SO(2)$ rotation in the $x$-$y$ plane by an angle $\vartheta$, as in eq.\eqref{rotmatrix}, can be undone by a global gauge transformation generated by the diagonal matrix $J_3$:
\bea
&& U(1)_C: \quad \langle A_j\rangle\,\to e^{i\vartheta J_3}\,\langle A_j\rangle\, e^{-i\vartheta J_3}\\\nonumber\\\nonumber
&& J_3\,\equiv\,{\rm diag}\left(-\tfrac{N-1}{2}, -\tfrac{N-3}{2}, -\tfrac{N-5}{2}\ldots , \tfrac{N-3}{2}, \tfrac{N-1}{2}\right)\,.
\eea
$J_3$ is the $N$-dimensional representation of one of the three generators of the $SU(2)$ algebra.
The phase rotation of the scalar VEV generated by $J_3$ can clearly be compensated by a $U(1)_B$ transformation. 

An interesting feature of the vacuum solution is that the Hermitean matrices $i\langle A_x\rangle$ and $i\langle A_y\rangle$ provide a matrix realization of coordinates on the noncommutative plane:
\be
\left[i\langle A_x\rangle,\, i\langle A_y\rangle\right]\,=\,2i\alpha^2 \left[\begin{array} {c|c}\quad{\mathds{1}}_{(N-1)\times(N-1)}\quad\quad & 0 \\
& \\
\hline
0\,& 1-\, N \end{array}\right]\,
\ee
where the noncommutativity parameter is $2\alpha^2$ as defined in eq.\eqref{alphabeta}, and scales as $\alpha^2\sim 1/N$ for large $N$
\footnote{ It is tempting to look for solutions to the vacuum equations which are reducible and consist of irreducible lower dimensional blocks each satisfying the finite dimensional algebra  implied by the vacuum conditions.   We have not succeeded in finding any solutions of this type.}. Furthermore, it appears that the coordinates are restricted to within a disc or droplet: 
\be
\left(i\langle A_x\rangle\right)^2\,+\,\left(i\langle A_y\rangle\right)^2\,=\,2\alpha^2{\rm diag}\left(1\,,\,3\,,5\,,\ldots\,, (2N-3)\,,\, (N-1)\right)\,.\label{droplet}
\ee
The radius of the droplet is bounded in the large $N$ limit since $\alpha^2 \sim 1/N$ with limiting value
\be
R_{\rm droplet}\left.\right|_{N\to\infty}\,=\,2\beta\sqrt{\frac{\mu_B}{\beta}-1}\,.
\ee
The algebra of matrices is closely related to that of harmonic oscillator creation and annihilation operators, when written in terms of the ladder operators:
\be
A^\pm\,=\, i\left(\langle A_x\rangle \,\pm\, i\langle A_y\rangle\right)\,,
\ee
which, for any finite $N$, satisfy $(A^+)^N\,=\,(A^-)^N\,=\,0$.
Precisely the same set of matrices were introduced to describe the  fractional quantum Hall droplet in \cite{Polychronakos:2001mi}, building on the connection between Abelian noncommutative Chern-Simons theory on the plane and the quantum Hall fluid \cite{Susskind:2001fb}. The matrix model has also been shown to describe the low energy dynamics of vortices in 2+1 dimensional Yang-Mills-Higgs theory with a Chern-Simons term \cite{Tong:2003vy, Tong:2015xaa}.
In this picture, the matrices $i\langle A_x \rangle$ and $i\langle A_y\rangle $ parametrize the (noncommuting) coordinates of $N$ particles in the droplet.  As eq.\eqref{droplet} indicates, the particles are placed in concentric circles of radius $\sim \sqrt{2n-1}$ for $n=0,1,2,\ldots, N-1$. In the present context, the two matrices appear to deconstruct two dimensions (at large $N$) on top of the 2+1 spacetime dimensions in which the field theory is originally formulated. 

Given the finite density ``droplet" ground state for general $N$, we need to calculate the spectrum of fluctuations around it. This will be addressed in detail in future work  \cite{ongoing}. However, we can already make a few remarks.  The spectrum must exhibit a massless state corresponding to the phonon arising from the spontaneous breaking of $U(1)_B$. In the droplet picture,  physical excitations live only on the boundary of the quantum Hall droplet and are associated to area preserving deformations of the droplet boundary, subject to a Gauss' law constraint following from the Chern-Simons equations of motion \cite{Polychronakos:2001mi}.  These have a zero mode corresponding to rotations of the circular droplet ground state, which could naturally be identified with the phonon.  In the language of the $N\times N$ matrices comprising the gauge field fluctuations, in an appropriate gauge (more precisely, unitary gauge), the excitations are encoded in the entries of the $N$-th row and column of gauge field fluctuations of ${\cal A}_x$ and ${\cal A}_y$, all other fluctuations corresponding to pure gauge or ``bulk" degrees of freedom  of the droplet. 
It would be extremely interesting to flesh out this picture in detail and explore the implications of this interpretation for the spectrum of the theory for generic $N$, and in particular its large-$N$ limit.
\section{Summary and future directions}
\label{sec5}
There are several immediate questions of interest that follow on from the results above.
The Bose condensed vacuum should have semiclassical vortex solutions, and it would be interesting to understand their explicit construction given the non-Abelian nature of the vacuum configuration. The ground state has a $U(1)$ colour-flavour locked global symmetry. A vortex solution that breaks this global symmetry will have an internal zero mode corresponding to a $U(1)$ moduli space of solutions.  Such vortices in a (non-Abelian) Higgs phase with noncommuting VEVs, carrying internal zero modes have been encountered previously in different contexts \cite{Markov:2004mj, Auzzi:2008ep, Auzzi:2009es}. The physical properties of such vortices and their role in the Bose-Fermi duality would be extremely interesting to explore.

The  origin of roton-like minima is often attributed to long range interactions. The interpretation of the background VEVs as noncommuting ``coordinates" for a quantum Hall droplet could thus provide a natural route to establish the existence of roton-like excitations\footnote{See e.g. \cite{Castorina:2004yd} for a discussion of the relation between noncommutative field theory and roton excitations in bosonic and fermionic systems.} for general $N>2$.  In general, the computation of the spectrum of excitations and their dispersion relations about the Bose condensed ground state should be facilitated by the connection to the droplet picture of \cite{Polychronakos:2001mi}. The goal would be to eventually understand the putative matching between the  spectra of the bosonic theory at weak 't Hooft coupling ($\lambda_B\ll 1$) and that of the dual critical fermion theory (coupled to Chern-Simons) at strong 't Hooft coupling $(\lambda_F\to 1)$. Perhaps the most puzzling aspect of this is the interpretation of the Higgsed ground state. When $\lambda_F=0$, and a $U(1)_B$ chemical potential is switched on in the critical theory, we do not  expect fermion bilinears to condense (see e.g. \cite{Hands:1998he}). As $\lambda_F$ is increased from zero it is conceivable that the effective potential for charged fermion bilinears carrying $U(1)_B$ favours a condensate either for any non-zero $\lambda_F$ or at some critical value. It would be extremely interesting to understand the behaviour of the large-$N$ effective potential for fermion bilinears for non-zero $\lambda_F$ and $\mu_B$.

 A related question has recently been explored in \cite{Choudhury:2018iwf} where Bose-Fermi duality at finite temperature and in the presence of scalar condensate has been established in the large-$N$ 't Hooft limit. We will need to understand  the modification of the zero temperature finite density state, and in particular the background gauge fields VEVs, by any non-zero temperature since the  Euclidean finite temperature theory is effectively two dimensional at long distances and thus fluctuations in the  phase of the scalar VEV are unsuppressed.  It will be interesting to understand the fate of the phonon-roton mode at finite temperature and non-zero 't Hooft coupling in the Chern-Simons-scalar theory.

\acknowledgments We would like to thank Justin David, Jeff Murugan and Carlos N\'u\~nez for  enjoyable discussions.
SPK acknowledges support from STFC grant ST/L000369/1. SS is supported by an STFC studentship under the DTP grant ST/N504464/1. DR is supported by a Royal Society SERB-Newton Fellowship.
\newpage

\appendix
\section{Determinant of fluctuation matrix for $SU(2)$ }
The determinant for the physical fluctuations is given in terms of frequency $\omega$ and momentum ${\bf p}$ as,
\be
\omega^8\,+\,\mu^2_BC_3\,\omega^6\,+\,\mu^4_BC_2 \,\omega^4 \,+\,\mu^6_BC_1\, \omega^2 \,+\,\mu^8_B\, C_0\,=\,0\,.
\ee
Assuming ${k>0, \mu_B>0}$ the coefficients $\{C_i\}$ are (for general choice of signs, it is understood that $k$ and $\mu_B$ will be replaced by their absolute values below):
\bea
&& C_0\,=\,\left(\frac{{\bf p}^2}{\mu_B^2}\right)^4\,+\,\left(\frac{{\bf p}^2}{\mu_B^2}\right)^3\left(\frac{m^2}{\mu_B^2}-1+\frac{6 g_4 v^2}{\mu_B^2}+\frac{17\pi^2 v^4}{k^2\mu_B^2}-\frac{12\pi v^2}{k\mu_B}\right)\,+\,\left(\frac{{\bf p}^2}{\mu_B^2}\right)^2\times\nonumber\\\nonumber\\\nonumber
&&\times\left(\frac{16 \pi ^2  m^2 v^4}{k^2 \mu_B^4}-\frac{12 \pi m^2 v^2}{k \mu^3_B}+\frac{96 \pi ^2 g_4 v^6}{k^2 \mu_B^4}-\frac{72 \pi g_4 v^4}{k \mu_B^3}+\frac{16 \pi^4  v^8}{k^4 \mu_B^4}-\frac{12 \pi ^3  v^6}{k^3 \mu_B^3}-\frac{16 \pi ^2
   v^4}{k^2 \mu^2_B}\right.\\\nonumber\\\nonumber
   &&\left.+\frac{12 \pi  v^2}{k \mu_B}\right)\,+\,\left(\frac{{\bf p}^2}{\mu_B^2}\right)\left(-\frac{384 \pi ^3 g_4 v^8}{k^3 \mu_B^5}+\frac{384 \pi ^2 g_4 v^6}{k^2 \mu_B^4}+\frac{960 \pi ^5
   v^{10}}{k^5 \mu_B^5}-\frac{1728 \pi ^4 v^8}{k^4 \mu_B^4}\right.\\\nonumber\\\label{ci}
   &&\left.-\frac{64 \pi ^3  m^2 v^6}{k^3 \mu_B^5}+\frac{832 \pi ^3
    v^6}{k^3 \mu_B^3}+\frac{64 \pi ^2 m^2 v^4}{k^2 \mu_B^4}-\frac{64 \pi ^2 v^4}{k^2 \mu_B^2}\right)\\\nonumber\\\nonumber
&&  C_1\,=\, -4\left(\frac{{\bf p}^2}{\mu_B^2}\right)^3\,+\,\left(\frac{{\bf p}^2}{\mu_B^2}\right)^2\left(-5-\frac{3m^2}{\mu_B^2}-\frac{18g_4v^2}{\mu_B^2}-\frac{51\pi^2 v^4}{k^2\mu_B^2}+\frac{28\pi v^2}{k\mu_B}\right)\,+\\\nonumber\\\nonumber
&&+\left(\frac{{\bf p}^2}{\mu_B^2}\right)\left(4 
   -4 \frac{m^2}{\mu_B ^2}-24 \frac{g_4  v^2}{\mu_B ^2}-\frac{192 \pi ^2 g_4 v^6}{k^2\mu_B ^4}+\frac{72 \pi   g_4 v^4}{k\mu_B ^3}-\frac{32 \pi ^4
  v^8}{k^4 \mu_B ^4}\right.\\\nonumber\\\nonumber
  &&\left.+\frac{76 \pi ^3    v^6}{k^3\mu_B ^3}-
  \frac{32 \pi ^2  m^2 v^4}{k^2\mu_B ^4}-\frac{84 \pi ^2
    v^4}{k^2\mu_B ^2}+\frac{12 \pi  m^2 v^2}{k\mu_B ^3}-\frac{28 \pi  v^2}{k\mu_B }\right)\,-\frac{384 \pi ^2g_4 v^6}{k^2 \mu_B ^4}\\\nonumber\\\nonumber
    &&-\frac{64 \pi ^4 v^8}{k^4 \mu_B ^4}+\frac{256 \pi ^3 v^6}{k^3 \mu_B
   ^3}-\frac{64 \pi ^2  m^2 v^4}{k^2 \mu_B^4}-\frac{192 \pi ^2v^4}{k^2 \mu_B^2}\\\nonumber\\\nonumber
   && C_2\,=\,6\left(\frac{{\bf p}^2}{\mu_B^2}\right)^2\,+\,\left(\frac{{\bf p}^2}{\mu_B^2}\right)\left(\frac{18  g_4 v^2}{\mu_B^2}+\frac{51 \pi ^2 v^4}{k^2 \mu_B^2}-\frac{20 \pi v^2}{k \mu_B }+\frac{3
   m^2}{\mu_B ^2}+13\right) \,+\frac{96 \pi ^2g_4 v^6}{k^2 \mu_B^4}
   \\\nonumber\\\nonumber
   &&+\frac{24  g_4 v^2}{\mu_B^2}
   +\frac{16 \pi ^4  v^8}{k^4 \mu_B^4}-\frac{64 \pi ^3 v^6}{k^3 \mu_B^3}+\frac{16 \pi ^2 m^2 v^4}{k^2 \mu_B^4}+\frac{116 \pi ^2 v^4}{k^2 \mu_B^2}-\frac{16 \pi  v^2}{k \mu_B }+\frac{4 m^2}{\mu_B^2}+12 \\\nonumber\\\nonumber
   && C_3\,=\, -4\left(\frac{{\bf p}^2}{\mu_B^2}\right)-\frac{6 g_B v^2}{\mu_B^2}-\frac{17 \pi ^2  v^4}{k^2 \mu_B^2}+\frac{4 \pi v^2}{k \mu_B
   }-\frac{m^2}{\mu_B^2}-7
\eea

\end{document}